\documentclass[11pt]{article}
\usepackage{epsfig,graphicx}

\usepackage{amsmath,amsfonts,amssymb} 
\usepackage{color}
\usepackage{cite}
\usepackage{slashed}
\usepackage[colorlinks = true,
            urlcolor  = blue,
            anchorcolor = blue]{hyperref}

\newcommand{\unit}{\leavevmode\hbox{\small1\kern-3.6pt\normalsize1}}

\def\locald{\rho_0}

\def\mnucl{m_{N}}
\def\redN{\mu_N}

\newcommand{\sigsinu}{\sigma^{SI,\,N}_0}
\newcommand{\sigsdnu}{\sigma^{SD,\,N}_0}

\def\lsim{\raise0.3ex\hbox{$\;<$\kern-0.75em\raise-1.1ex\hbox{$\sim\;$}}}
\def\gsim{\raise0.3ex\hbox{$\;>$\kern-0.75em\raise-1.1ex\hbox{$\sim\;$}}}

\allowdisplaybreaks

\begin{document}

\thispagestyle{empty}
\begin{flushright}
  FTUAM-15-21\\
  IFT-UAM/CSIC-15-080\\

 % \vspace*{2.mm}{\color{red}{Modified: \today}}
\end{flushright}

\begin{center}
  {\Large \textbf{ On the importance of direct detection combined limits for spin independent and spin dependent dark matter interactions
  } }  
  
  \vspace{0.5cm}
  Cristina Marcos,
  Miguel Peir\'o and Sandra Robles\\[0.2cm] 
    
  {\textit{  
      Instituto de F\'{\i}sica Te\'{o}rica
      UAM/CSIC \& Departamento de F\'{\i}sica Te\'{o}rica,
      Universidad Aut\'{o}noma de Madrid, 28049
      Madrid, Spain\\[0pt] 
  }}

\vspace*{0.7cm}

\begin{abstract}

In this work we show how the inclusion of dark matter (DM) direct detection upper bounds in a theoretically consistent manner can affect the allowed parameter space of a DM model. Traditionally, the limits from DM direct detection experiments on the elastic scattering cross section of DM particles as a function of their mass are extracted under simplifying assumptions. Relaxing the assumptions related to the DM particle nature, such as the neutron to proton ratio of the interactions, or the possibility of having similar contributions from the spin independent (SI) and spin dependent (SD) interactions can vary significantly the upper limits. Furthermore, it is known that astrophysical and nuclear uncertainties can also affect the upper bounds. To exemplify the impact of properly including all these factors, we have analysed two well motivated and popular DM scenarios: neutralinos in the NMSSM and a $Z'$ portal with Dirac DM. We have found that the allowed parameter space of these models is subject to important variations when one includes both the SI and SD interactions at the same time, realistic neutron to proton ratios, as well as using different self-consistent speed distributions corresponding to popular DM halo density profiles, and distinct SD structure functions. Finally, we provide all the necessary information to include the upper bounds of SuperCDMS and LUX taking into account all these subtleties in the investigation of any particle physics model. The data for each experiment and example codes are available at this site \url{http://goo.gl/1CDFYi}, and their use is detailed in the appendices of this work.

\end{abstract}

\end{center}

	\newpage
	
% -----------------------------------------------------------
% ARTICLE
% -----------------------------------------------------------

%\tableofcontents

\section{Introduction}

Nowadays, the dark matter (DM) nature is one of the greatest mysteries of modern physics. The possibility of non gravitational interactions of DM with ordinary matter is one of the key pieces to understand the DM problem. In this sense, the experimental community is putting a great deal of effort to provide such a valuable information.

Direct searches for DM aim at detecting the scattering of DM particles off nuclei inside a low-background target detector. Once a DM particle hits a nucleus, the recoiling nucleus will release a certain amount of energy that, above a given threshold, can be measured using different and sophisticated techniques. This field is currently undergoing an exciting period, with several experimental collaborations, using different targets and techniques, reporting potential signals of DM that are, nevertheless, in conflict with null results in other experiments. Most remarkably, the results obtained by the SuperCDMS~\cite{Agnese:2014aze} and LUX~\cite{Akerib:2013tjd} Collaborations have placed the most stringent upper bounds on the elastic scattering spin independent (SI) cross section up to date, for DM masses above 4 GeV. Particle physics models for DM are then tested using these results, which in general, reduce considerably the allowed region of the parameter space of a given model. 

We are living in the precision data era. There is a huge amount of experimental data from colliders, direct detection and indirect detection experiments, that must be taken into account to determine the viability of a model of physics beyond the Standard Model (BSM). In this regard, theoreticians are making a special effort to reduce the uncertainties from the theoretical side, for instance, in the calculation of the Higgs mass, which improves notably the interpretation of the experimental data in the context of different BSM scenarios.

Direct detection experimental collaborations adopt a set of assumptions about the DM nature and the DM halo which are useful in order to compare different results within a unified framework. More specifically, their results are traditionally presented considering only SI interactions\footnote{Sometimes, they also present bounds considering only spin dependent (SD) interactions under certain assumptions about the neutron to proton ratio $a_n/a_p$.} and the same interaction strength of DM with protons and neutrons, i.e. with a neutron to proton ratio $f_n/f_p=1$.  The speed distribution of DM particles is taken to follow a Maxwell-Boltzmann distribution, a consequence of assuming the halo to be an isothermal and isotropic sphere, the Standard Halo Model (SHM). Then, these results are used to constrain the parameter space of models, like supersymmetry (SUSY), in which theoretical calculations are performed with a high degree of accuracy. Furthermore, in many models the assumptions that experimental collaborations consider do not apply neither for the DM particle considered nor for the DM halo profile\footnote{For DM candidates with a mass heavier than approximately 100 GeV the impact of speed distribution uncertainties on the upper limits of the elastic scattering cross section can be generally neglected.}. This is a consequence of the aforementioned over-simplifying assumptions, as realistic particle physics models of DM (e.g. SUSY) allow for different couplings to neutrons and protons and similar contributions from SI and SD interactions. Then, in principle, it is not fully consistent to employ the upper limits quoted by experimental collaborations to constrain general particle physics models of DM. However, to take into account all these subtleties is not an easy task, since it would require the simulation of a direct detection experiment, which makes very difficult their inclusion when scanning the parameter space of a model.

All in all, it is usual to compare the predictions of a given model with direct detection upper limits at face value, this is, considering the SI and SD interactions separately\footnote{Sometimes, not even considering the SD interactions for a spin $1/2$ DM particle.}, an equal contribution from protons and neutrons, i.e. $f_n/f_p=1$, and assuming the SHM. Nonetheless, it has been pointed out that either the neutron to proton ratios~\cite{Kurylov:2003ra,Giuliani:2005my,Chang:2010yk,Feng:2011vu,Feng:2013fyw,Cirigliano:2013zta} or the SD structure functions~\cite{Cerdeno:2012ix} and speed distributions~\cite{Belli:2002yt,Strigari:2009zb,McCabe:2010zh,Frandsen:2011gi,Green:2011bv,Fairbairn:2012zs,Bernal:2014mmt} chosen can vary the interpretation of direct detection data substantially. In fact, to account for the latter effect, several halo independent methods have been proposed recently
~\cite{Gondolo:2012rs,DelNobile:2013cta,Bozorgnia:2013hsa,DelNobile:2013cva,Fox:2014kua,Feldstein:2014gza,Scopel:2014kba,Cherry:2014wia,Feldstein:2014ufa,Anderson:2015xaa,Ferrer:2015bta,Gelmini:2015voa}.
In this article, we present a set of tabulated data that contain the necessary information to calculate the expected number of DM induced events in the SuperCDMS and LUX experiments taking into account different possibilities of the mentioned ingredients. These data allow a fast check of the validity of a given model solution considering both, the SI and SD interactions, for arbitrary values of $f_n/f_p$ and $a_n/a_p$, using realistic DM speed distributions and including the uncertainties related to the SD form factors. We provide the data and example codes at this site \url{http://goo.gl/1CDFYi}.

To this end, we have carried out a decomposition of the SuperCDMS and LUX detector signals in intensity (the part related to the cross sections) and shape (the part dependent on the energy) in such a way that the total number of predicted events in these experiments can be easily calculated even if the particle physics and/or DM profile assumptions made by collaborations do not allow a consistent and serious test. The tabulated data we have derived are very similar to the \textit{scaling functions} introduced in Ref.~\cite{DelNobile:2013sia}\footnote{Notice that the definition of these functions in Ref.~\cite{DelNobile:2013sia} varies slightly respect to the case presented here.}. Here, we extend this information to include more realistic descriptions of the DM halo profile, and different SD structure functions.
These data can be used to evaluate at which confidence level a certain point of the parameter space of a model is excluded or allowed, without any computationally intensive calculation. Such an observable can then be translated into upper limits on the properties of the DM candidate. The aim behind this work is to provide the particle physics community with enough information to include direct detection bounds in terms of the DM expected number of events, in a complete and consistent manner.

To exemplify the impact that the combined limits (using the SI and SD interactions at the same time), together with distinct assumptions about the speed distribution
and the SD structure functions, have on the allowed solutions in a model parameter space, we have chosen as case studies two well motivated and popular models: the Next-to-Minimal Supersymmetric Standard Model (NMSSM) and a $Z'$ portal with Dirac DM. We show explicitly step by step how the previous considerations can affect the phenomenological viability of the solutions found. Moreover, the data provided in this work for different DM halo profiles, can be used to perform a consistent analysis of a model in light of DM direct and indirect detection experiments using the same DM halo density profile assumptions.

This paper is organised as follows. In Section~\ref{sec:basics}, we present the basic expressions that allow to compute a DM signal in the LUX and SuperCDMS detectors. Then, in Section~\ref{sec:upper} we derive the upper limits from these experiments under different assumptions in the velocity distribution and SD form factors, emphasizing their impact on the predicted number of events. In Section~\ref{sec:nmssm} and \ref{sec:zprime}, we show how the inclusion of the upper bounds in terms of the number of events, considering both the SI and SD interactions, realistic neutron to proton ratios and different speed distributions, has a notable impact on the allowed parameter space of the NMSSM and $Z'$ portal models, respectively. Then, we repeat this exercise using different SD structure functions, in such a way that we can quantify the effect on the parameter space of these scenarios. Finally, the conclusions are left for Section~\ref{sec:conclusions}. In appendix~\ref{sec:functions}, we give detailed information about the extraction of the upper bounds and how to apply them to a generic model, and in appendix~\ref{sec:files} we briefly describe the data files and example codes attached to this work.

\section{Direct detection of dark matter}
\label{sec:basics}

Let us start by summarising the basic expressions that describe the differential rate in a DM direct detection experiment \cite{Smith:1988kw} (for a recent review, see Ref.\,\cite{Cerdeno:2010jj}).
The differential event rate for the elastic scattering of a WIMP with mass $m_{\chi}$ off a nucleus with mass $\mnucl$ reads\footnote{ Typically given in units of counts/day/kg/keV.} 
\begin{equation}
  \frac{dR}{dE_R}=\frac{\locald}{\mnucl\,m_{\chi}}\int_{v_{min}}^{v_{esc}}  v
  f(v)\, \frac{d\sigma}{dE_R}(v,E_R)\, d v\,,
  \label{drate}
\end{equation}
where $\locald$ is the local DM density, $f(v)$ is the DM speed distribution normalised to unity and velocities are expressed in the detector reference frame. The integration over the DM velocity is computed from the minimum DM speed needed to induce a recoil of energy $E_{R}$, $v_{min}=\sqrt{(\mnucl E_R)/(2\redN^2)}$, where $\redN$ is the WIMP-nucleus reduced mass, to the  escape velocity $v_{ esc}$ also in the detector reference frame, above which DM particles are not gravitationally bounded to the Galaxy.
The DM-nucleus differential cross section $d\sigma/dE_R$ is computed from the Lagrangian that describes the interaction of a given WIMP with ordinary matter and encodes the microscopic interactions of DM particles and quarks inside nucleons.

As mentioned before, direct detection experiments usually assume the SHM, which describes the Galaxy as an isotropic isothermal sphere. This model provides a useful and common framework to compare different experiments, but definitely it is not a realistic description of our Galaxy. Due to this fact, we will use the mean speed distributions of Ref.~\cite{Fornasa:2013iaa} derived separately for three possible choices of the DM density profile, namely a Navarro-Frenk-White (NFW) profile, an Einasto one and a Burkert profile\footnote{The mean values of the escape velocity, Sun's velocity, and local DM density for each profile can be found in tables 2, 3 and 4 of Ref.~\cite{Fornasa:2013iaa}.}. These distributions are extracted from self-consistent DM phase space distributions that are in agreement with the astrophysical observables of the Galaxy and thereby, we expect them to give a more realistic description of the speed distribution of DM particles in the Solar neighbourhood. For completeness and to make it easy for the reader to compare with other results in the literature, including those provided by the experimental collaborations, we will use the SHM as well. 

In general, the DM-nucleus scattering cross section can be split into a SI and a SD contribution. The total cross section can be then calculated by adding these terms coherently, using nuclear wave functions. Thus, the differential cross section is given by
\begin{equation}
	\frac{d\sigma}{dE_R} = \frac{m_N}{2 \mu_N^2 v^2}
	\left(\sigsinu F^2_{SI}(E_R) + 
	\sigsdnu F^2_{SD}(E_R) \right),
	\label{eqn:SI_and_SD}
\end{equation}
where $\sigsinu$ and $\sigsdnu$ are the SI and SD components of the cross sections at zero momentum transfer, and the form factors $F^2_{SI,\,SD}(E_R)$ account for the coherence loss which leads to a suppression in the event rate for heavy nuclei in the SI and SD contributions (see Ref.~\cite{Lewin:1995rx} for a complete description of these prescriptions). Thanks to the Fermi's Golden Rule in the Born approximation, we can factorize out all the energy (momentum) dependence of the scattering of DM off a nucleus inside a form factor $F^2(q)$. This allows us to express the zero momentum transfer cross sections as~\cite{Lewin:1995rx,Jungman:1995df}
\begin{eqnarray}
\sigsinu &=& \frac{\redN^2}{\mu^2_p}\,\left[Z +(A-Z)\frac{f_n}{f_p}\right]^2\sigma^{SI}_p\,,\nonumber\\
\sigsdnu &=&  \frac{\redN^2}{\mu^2_p}\,\left[S_p+ S_n\frac{a_n}{a_p}\right]^2\,\left(\frac{4}{3}\frac{J+1}{J}\right)\sigma^{SD}_p\,,
\label{eq:zeromoment}
\end{eqnarray}
where $\mu_p$ is the proton-WIMP reduced mass; $S_p$ and $S_n$ are the expectation values of the total proton and neutron spin operators; $f_p$, $f_n$ and $a_p$, $a_n$ are the effective WIMP couplings to protons and neutrons in the SI and SD case, respectively\footnote{Notice that for simplicity we have not included here a possible vector coupling (corresponding to non-Majorana DM particles), which would lead to an extra contribution in the expression for $\sigsinu$. }. The target material is defined by the following parameters, the atomic number $Z$, the mass number $A$, and the total nuclear spin $J$. In these expressions, we have defined the WIMP-proton cross sections ($\sigma^{SI}_p$,$\sigma^{SD}_p$) as,
\begin{eqnarray}
\sigma^{SI}_p &=& \frac{4}{\pi}\mu^2_pf_p^2\nonumber\\
\sigma^{SD}_p &=&  \frac{24G_F^2}{\pi}\mu^2_pa_p^2\,,
\label{eq:fp_general}
\end{eqnarray}
where $G_F$ is the Fermi coupling constant and the spin of the DM particles is considered equal to that of the proton.  Note that Eq. (\ref{eq:zeromoment}) reduces to these expressions when the quantities referred to the nucleus are substituted by those of the proton, i.e. using $S_p=1/2$, $S_n=0$, $J=1/2$, $A=Z=1$.

\subsection{Differential rate in SuperCDMS}

The actual energy observed by the SuperCDMS experiment is not the recoil energy presented in Eq.\eqref{drate}, but the total phonon energy produced by a collision event. To take into account such a change, we follow the prescription given in Ref.~\cite{supercdms}. The phonon energy, $E_p$, can be expressed as the sum of the recoil energy and the Neganov-Luke phonon energy, produced proportionally to the charge energy,
\begin{equation}
E_{p} = E_{R}+\frac{E_Q}{3 eV}e\Delta V.
\end{equation}
where $\Delta V=4 V$ is the bias voltage applied to the detectors, $e$ is the electron charge, and finally $E_Q$ is the mean charge for nuclear recoils. The latter is calibrated as a function of $E_p$ and then parametrised as follows,
\begin{equation}
E_Q=f(E_p)=\alpha_1+\alpha_2E_p+10^{\alpha_3}{\rm erf}\left(-\frac{E_p}{10^{\alpha_4}}\right),
\end{equation}
where the parameters $\alpha_i$ depend on the corresponding detector and can be found in Ref.~\cite{supercdms}. Finally, to convert the differential rate terms of the nuclear recoil energy, given in Eq.\eqref{drate}, to the differential rate in total phonon energy, we apply the following change of variables,
\begin{eqnarray}
\frac{dR}{dE_p}(E_p)&=& \frac{dR}{dE_R}(E_R(E_p))\times\frac{dE_R}{dE_p}(E_p),\nonumber\\
\frac{dE_R}{dE_p}(E_p) &=&  1-f'(E_p)\times \frac{\Delta V}{3 V}\,.
\label{eq:scdms-drde}
\end{eqnarray}

\begin{figure}[t!]
	\begin{center}  
	\epsfig{file=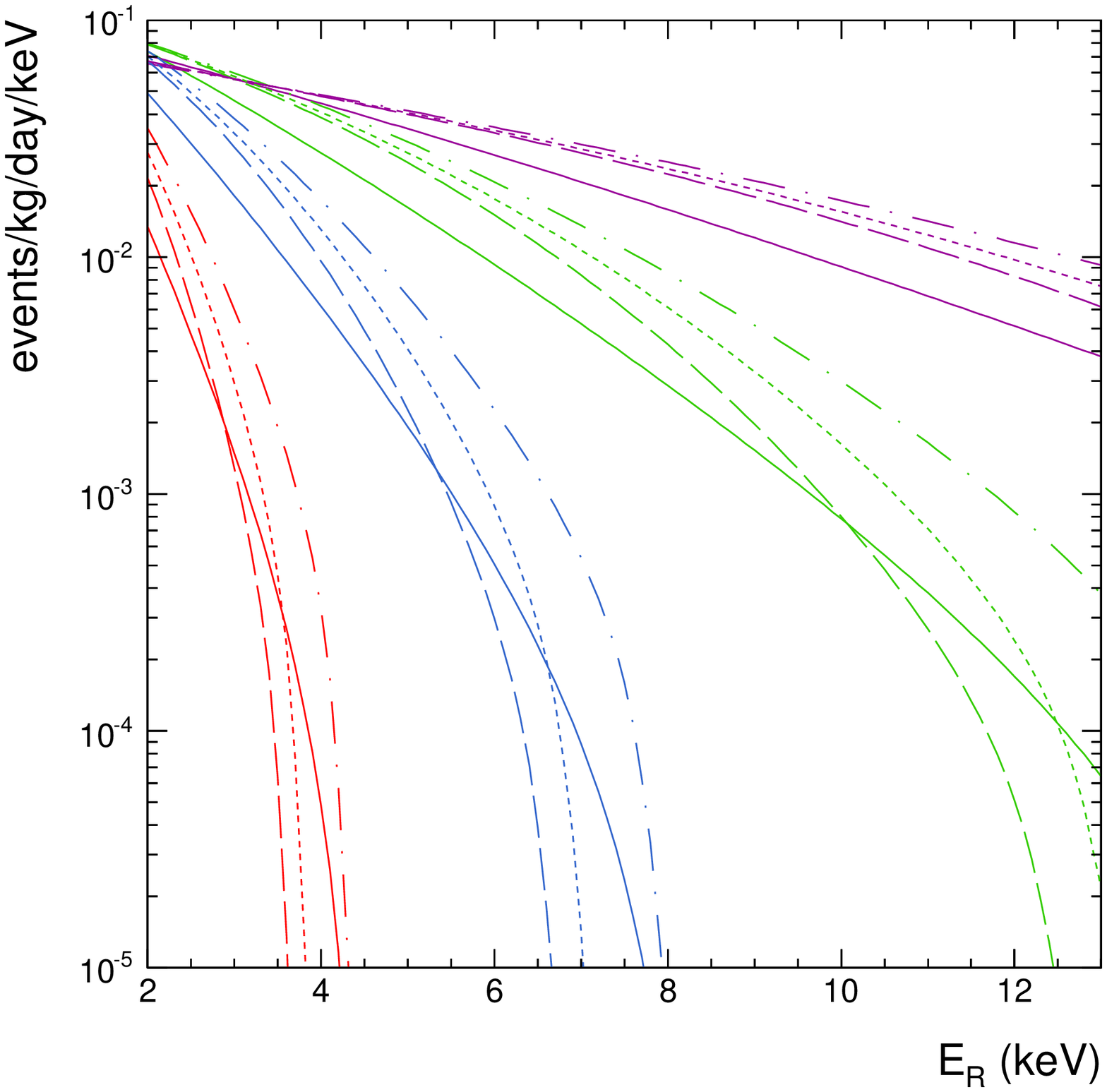,width=7.6cm}
	\epsfig{file=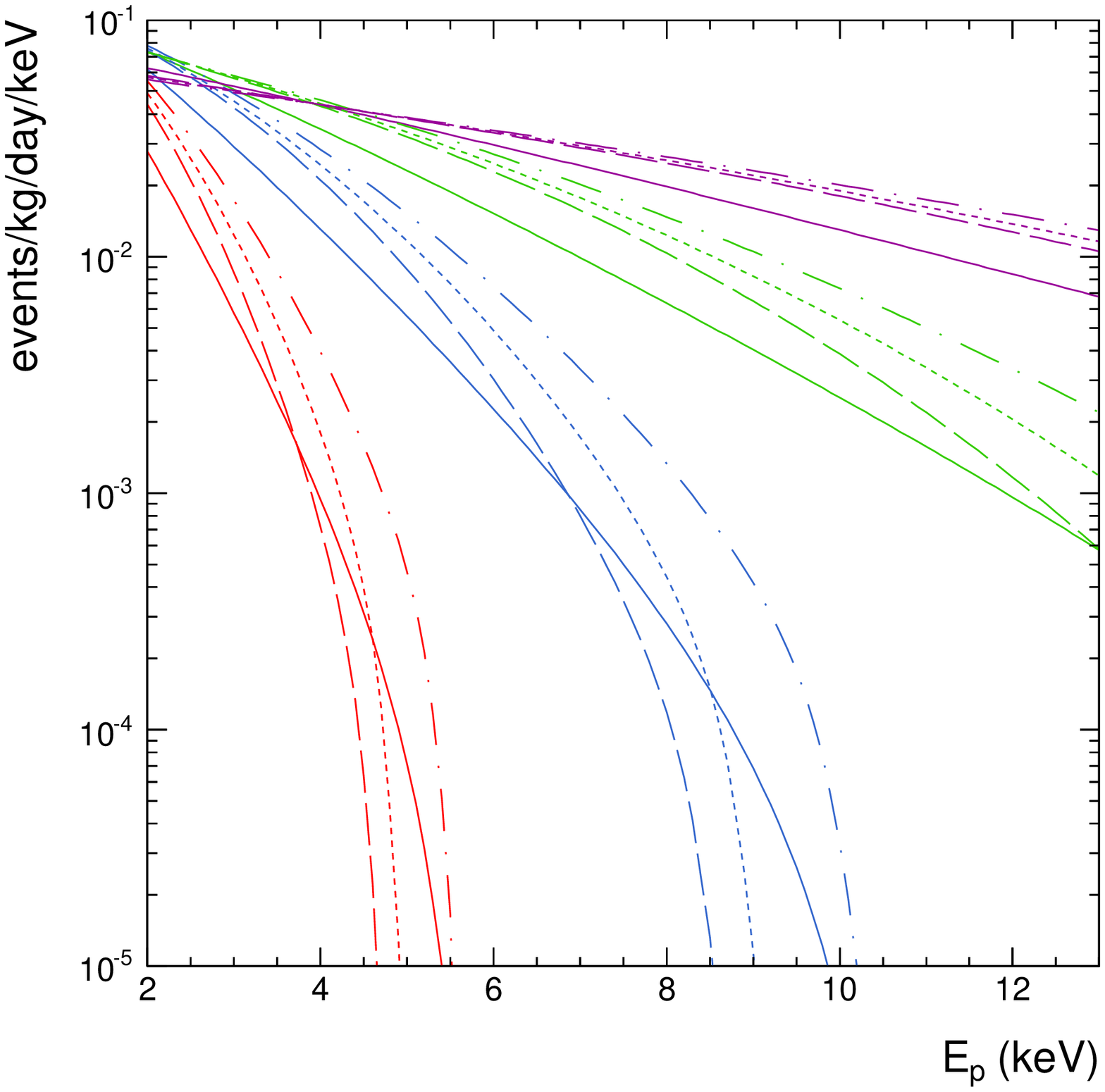,width=7.6cm}
	\end{center}
\caption{\small Differential spectrum for DM-nucleon scattering on Ge in terms of the recoil energy (left) and the total phonon energy for the T2Z1 charge model (right) for different DM masses: $m_{\chi}=5$~GeV (red), $m_{\chi}=7$~GeV (blue), $m_{\chi}=10$~GeV (green) and $m_{\chi}=15$~GeV (magenta). The SI cross section for protons and neutrons is $\sigma^{SI}_{p,n}= 10^{-6}$~pb, while the SD cross sections are set to zero. Different assumptions about the DM speed distribution are used: SHM (solid) and self-consistent speed distributions corresponding to a NFW (dashed-dotted), Einasto (dotted) and Burkert DM profiles (dashed) from Ref.~\cite{Fornasa:2013iaa}.
}
  \label{fig:rates_Ge}
\end{figure}

In figure~\ref{fig:rates_Ge}, we show the differential rate in a Ge experiment as a function of the recoil energy (left panel) and the total phonon energy (right panel) for a SI cross section off nucleons $\sigma^{SI}_{p,n}= 10^{-6}$~pb, neglecting contributions from the SD interactions. In the right panel, we have used the charge model of the T2Z1 detector as it is given in Ref.~\cite{supercdms}. We have generated the differential rates for distinct DM masses and different speed distributions to show explicitly the impact that the velocity distribution choice has in the differential rate. Our SHM results (solid lines) are in very good agreement with that of the SuperCDMS Collaboration when comparing these results with figure 2 of Ref.~\cite{supercdms}. As it is shown, generally, the variation of the speed distribution tends to predict higher rates in the low mass region (all DM masses used in this figure would fall in such category) with respect to the SHM case. This would induce more stringent bounds as we will see later on.

\subsection{Differential rate in LUX}

\begin{figure}[t!]
	\begin{center}  
	\epsfig{file=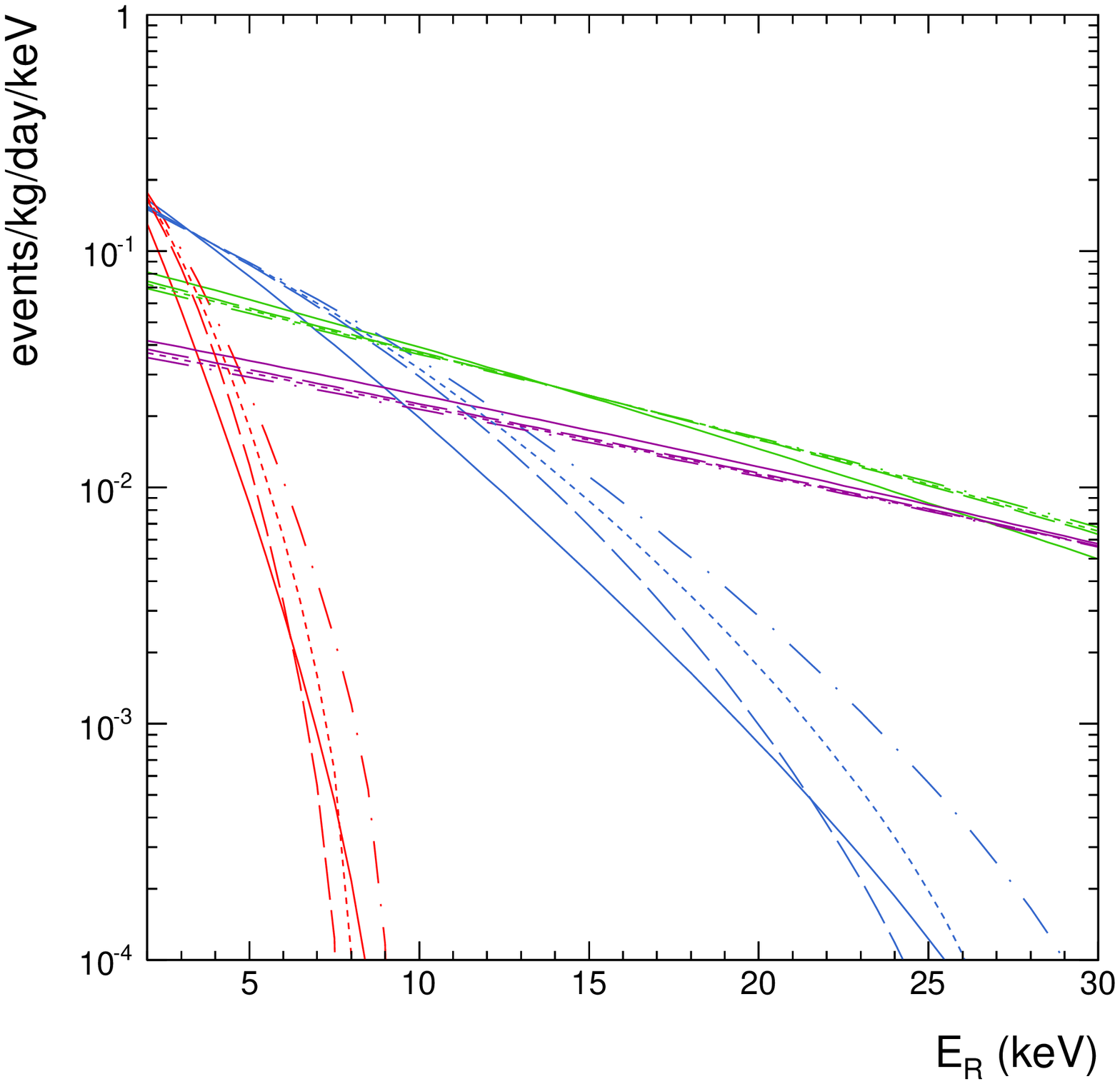,width=7.6cm}
	\epsfig{file=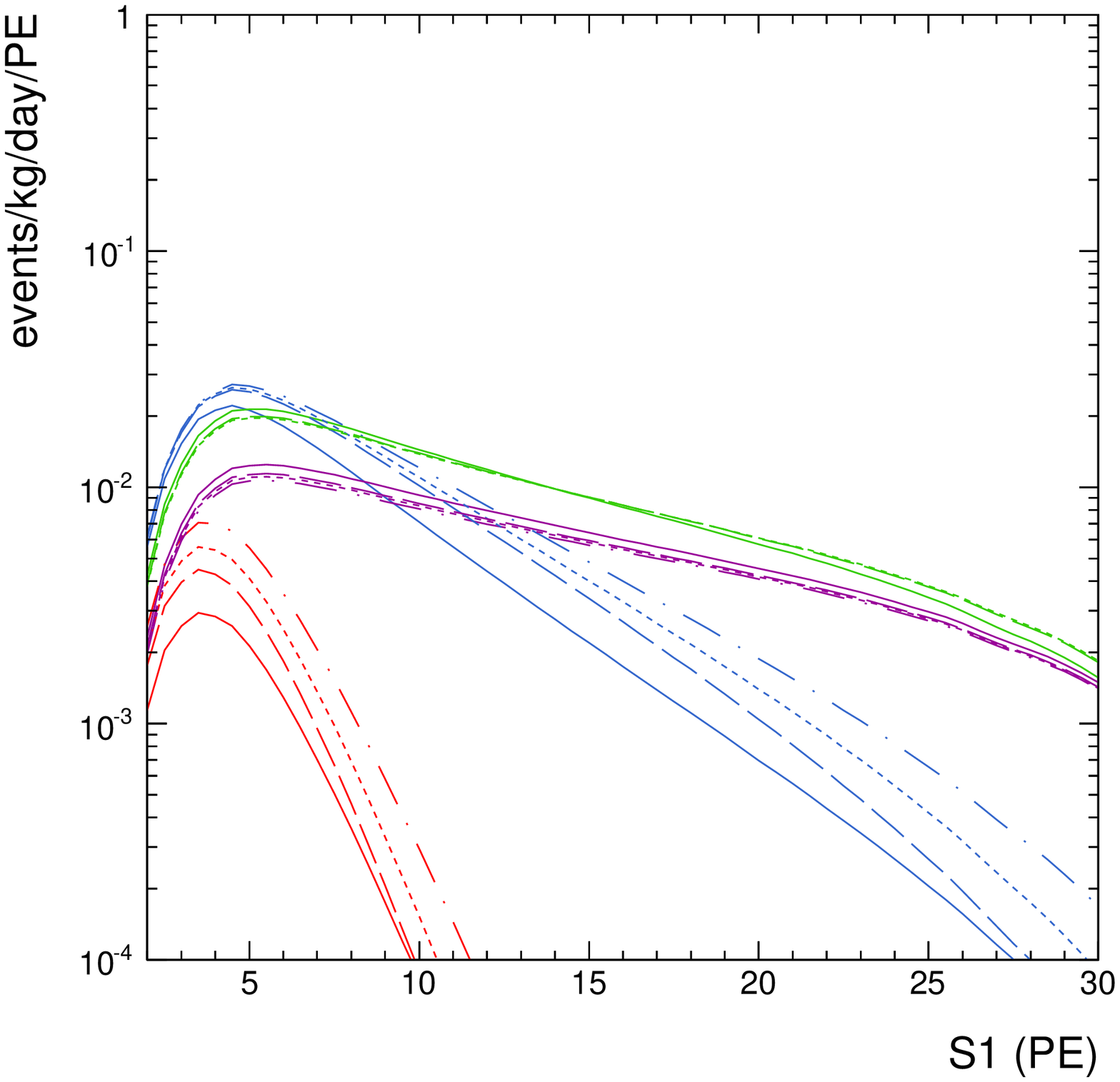,width=7.6cm}
	\end{center}
\caption{\small Differential spectrum for WIMP-nucleon scattering on Xe in terms of the recoil energy (left) and the scintillation signal $S1$ in PE (right) for different WIMP masses: $m_{\chi}=10$~GeV (red), $m_{\chi}=20$~GeV (blue), $m_{\chi}=50$~GeV (green) and $m_{\chi}=100$~GeV (magenta). The cross sections and speed distributions used are the same as in Figure~\ref{fig:rates_Ge}. In the right panel, we have included the efficiency of the LUX detector.
}
  \label{fig:rates_Xe}
\end{figure}

We address now the LUX experiment. In general, the observed energy in liquid Xe-based detectors is the number of photoelectrons~(PE) $S1$. To simulate the signal in LUX, we have followed the description given in Ref.~\cite{Aprile:2011hx}, where the number of expected PE, $\nu(E_R)$, reads
\begin{equation}
\nu(E_R) = E_{R} \mathcal{L}_{eff}(E_{R}) Q_{\gamma},
\end{equation}
where $Q_{\gamma}=0.14$ is the photon detection efficiency for LUX~\cite{Akerib:2013tjd}. $\mathcal{L}_{eff}(E_{R})$ is the absolute scintillation yield, which we have extracted from Ref.~\cite{lux-leff}. The signal rate in number of photoelectrons $n$ is then given by
\begin{equation}
\frac{dR}{dn} = \int_0^\infty dE_R \frac{dR}{dE_R}\times {\rm Poiss}\left( n | \nu(E_R) \right),
\end{equation}
where ${\rm Poiss}\left( n | \nu(E_R) \right)$ is a Poisson distribution with expectation value $\nu(E_R)$, and the differential rate in $E_R$ is given by Eq.~\eqref{drate}.
Taking into account the finite average single-photoelectron resolution $\sigma_{\rm PMT}=0.37$~PE of the photomultipliers~\cite{pmt}, the resulting $S1$-spectrum is given by
\begin{equation}\label{eq:rate}
\frac{d R}{dS1} =\sum_{n=1}^\infty {\rm Gauss}( S1 | n ,\sqrt{n}\sigma_{\rm PMT} ) \times \frac{dR}{dn} \times\zeta(S1),
\end{equation}
where $\zeta(S1)$ is the acceptance of the applied cuts. $ {\rm Gauss}( S1 | n ,\sqrt{n}\sigma_{\rm PMT} )$ denotes a normal distribution with mean $n$ and standard deviation $\sqrt{n}\sigma_{\rm PMT}$.

In figure~\ref{fig:rates_Xe}, we show a set of differential rates for different DM masses and speed distributions, as a function of the recoil energy (left panel) and $S1$ (right panel) in the DM search energy window. The non-monotonic behaviour of the differential rates in the right panel is due to the efficiency of the LUX detector. As we can note, now the spectra are flatter than for Ge (see figure~\ref{fig:rates_Ge}) as a consequence of the heavier Xe isotopes. The same tendency as before is observed, the predicted rates are higher for the self-consistent speed distributions than that predicted for the SHM. In this case, since the DM masses used here are heavier in some cases than that considered for SuperCDMS, we see that for a DM of 50 GeV the effect of the speed distribution starts to be negligible. This effect will also appear in the LUX upper limits.

\section{Upper limits for SuperCDMS and LUX}
\label{sec:upper}

To extract the upper limits, we use the Yellin's maximum gap method~\cite{Yellin:2002xd} at 90\% C.L., which has been shown to produce good results for both LUX~\cite{Bozorgnia:2013pua,Gresham:2013mua,DelNobile:2013gba,Fan:2013bea,Fox:2013pia,Gelmini:2014psa,DelNobile:2014eta,DelNobile:2014sja,Bozorgnia:2014gsa} and SuperCDMS~\cite{DelNobile:2014eta,Gelmini:2014psa,DelNobile:2014sja,Bozorgnia:2014gsa}. This statistical tool allows to extract the upper bounds in presence of an unknown background in terms of the expected DM events in the signal region. To such end, we calculate the total number of expected DM events using the procedure outlined in Appendix~\ref{sec:functions} for different speed distributions as well as SD structure functions. Notice that this method does not permit to combine the information from both experiments at the same time, however, there has been other proposed methods, like the ones presented in Refs.~\cite{Bozorgnia:2014gsa,Feldstein:2014ufa}, that allow such a combination. In the following, we provide the information about the SuperCDMS and LUX experimental setups used to calculate the tabulated data attached to this work, and therefore, the upper limits.

\noindent $\bullet$ \textbf{SuperCDMS:} The SuperCDMS Collaboration has reported a first search for DM using the latest background
rejection capabilities for an exposure of 577 kg-days~\cite{Agnese:2014aze}. This analysis relies on a low-threshold performance of the detector, and hence, it was carried out only for DM masses below 30 GeV. Eleven events were observed after unblinding, allowing to set the most stringent upper bound on the SI DM-nucleon cross section for masses approximately below 6 GeV. For this analysis, only seven detectors (T1Z1, T2Z1, T2Z2, T4Z2, T4Z3, T5Z2 and T5Z3) had significant sensitivity to very low energy recoils, and then these were employed to set the upper limit. The remaining detectors were used to veto events that scatter in multiple detectors instead, since multiple scatter event topology has a negligible probability to occur for DM.

\begin{figure}[t!]
	\begin{center} 
	\epsfig{file=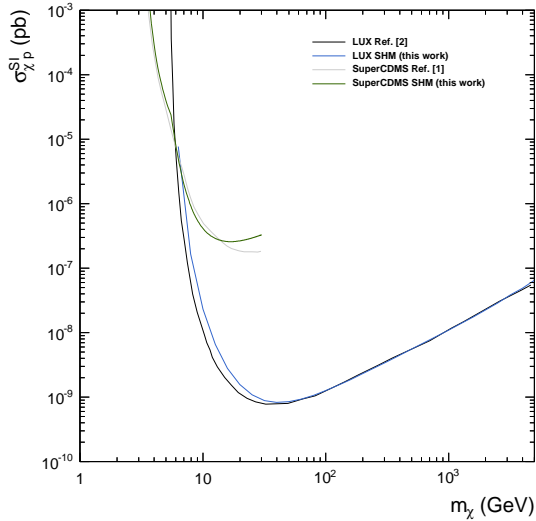,width=7.6cm} 
	\end{center}
\caption{\small Upper bounds on the SI proton cross section as a function of the DM mass. Black and grey curves are the official upper bounds from the LUX and SuperCDMS Collaborations, respectively. Blue and green curves are our calculations of the 90\% C.L. upper limits for the LUX and SuperCDMS experiments respectively using the Yellin's maximum gap method.  
}
  \label{fig:si-m-comparison}
\end{figure}

We calculate the expected signal for each detector separately using the efficiencies, exposures and charge models specified in Ref.~\cite{supercdms} in terms of the total phonon energy given in Eq.~\eqref{eq:scdms-drde}. We also include an estimation of the energy resolution, considering the 1.3 keV activation line at a total phonon energy of approximately 3 keV from Fig. 3 of Ref.~\cite{Agnese:2014aze}, to be energy independent and $\Delta E_p=0.25$ keV. To find the upper limits using the maximum gap method we have taken into account the eleven observed events in the signal region $E_p=$[2-12.5]~keV.

\noindent $\bullet$ \textbf{LUX:} The LUX Collaboration reported null results in the search for DM particles for an exposure of 10065 kg days, which allowed them to place the most stringent upper limits on the SI interactions of DM off nucleons above 6 GeV approximately. The signal region was set to be 2-30 PE in $S1$. The acceptance of the detector is shown in the bottom of Fig. 1 of Ref.~\cite{Akerib:2013tjd} and we add an extra $1/2$ factor to account for the 50\% of nuclear recoil acceptance. We use the $S1$ single PE resolution $\sigma_{PMT} = 0.37$ PE~\cite{pmt}, a 14\% of photon detection efficiency, and the absolute scintillation efficiency digitized from Ref.~\cite{Akerib:2013tjd}. 

In figure~\ref{fig:si-m-comparison}, we show the upper bounds on the SI proton cross section versus the DM mass for SuperCDMS and LUX experiments. In black (grey), we have plotted the official limit of the  LUX~\cite{Akerib:2013tjd} (SuperCDMS~\cite{Agnese:2014aze}) Collaboration, whereas in blue and green, respectively, we represent our findings of the upper limits at 90\% C.L. using the Yellin's maximum gap method. For consistency, our results have been obtained using the same speed distributions as in the case of the collaborations. In the case of LUX, to extract the upper bounds we have considered zero candidate events observed, even though the results of the collaboration showed one candidate event marginally close to the background region in the $\log_{10}(S2/S1)-S1$ plane. This fact is reflected as a slight disagreement between our results and those from the collaboration for masses below 30 GeV. On the contrary, for heavier masses, the agreement between our findings and the official limit is excellent. For SuperCDMS, the small discrepancies between our results and those of the collaboration are only due to the statistical method applied, since the latter used the Yellin's optimum interval method~\cite{Yellin:2002xd}.

\subsection{Limits on the SI cross section}

\begin{figure}[t!]
	\begin{center} 
	\epsfig{file=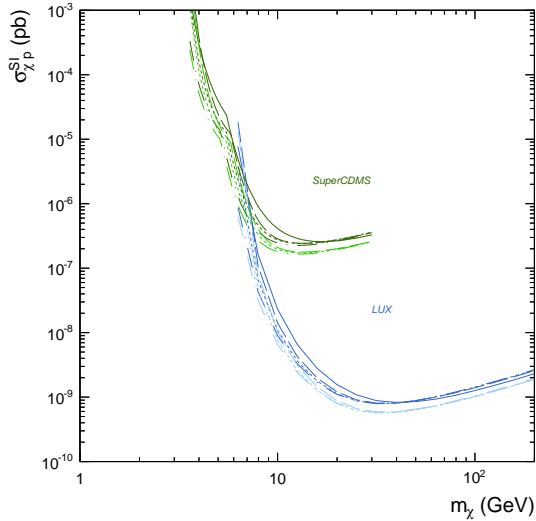,width=7.6cm}
	\end{center}
\caption{\small Upper bounds on the SI proton cross section as a function of the DM mass for the LUX (blue curves) and SuperCDMS (green curves) experiments. The DM speed distribution used in each case is: SHM (solid), NFW (dashed-dotted), Einasto (dotted) and Burkert (dashed). Darker colours correspond to $\rho_0=0.3$~GeV/cm$^3$ while lighter colours to the mean value of the local density extracted from Ref.~\cite{Fornasa:2013iaa}.
}
  \label{fig:si-m}
\end{figure}

Let us start with the results for the SI interactions of DM off nucleons. In figure~\ref{fig:si-m}, we show the upper limits for the LUX (blue) and SuperCDMS (green) experiments. All these curves have been derived under the usual assumptions, $f_n/f_p=1$ and $\sigma^{SD}_{p,n}=0$. Different type of curves correspond to different choices of the DM speed distribution. For the SHM (solid curve), we have used the same values of the speed distribution as in Refs.~\cite{Agnese:2014aze,Akerib:2013tjd} with a local DM density $\rho_0=0.3$~GeV/cm$^3$. On the other hand, for the remaining profiles, our results have been obtained with the speed distributions from Ref.~\cite{Fornasa:2013iaa} and are shown as dashed-dotted curves for the NFW profile, dotted for the Einasto profile, and dashed for the Burkert profile. In addition, we have rescaled all the upper bounds according to different values of the local DM density. We depict in darker colours the upper limits in which we have used the same value of the local density as in the SHM case, in order to show how the shape of the DM velocity distribution impacts on the upper bounds. In contrast, curves with lighter colours correspond to upper limits derived using the mean values that the authors of Ref.~\cite{Fornasa:2013iaa} obtained for each of the above mentioned profiles, namely, $\rho^{NFW}_0=0.41$, $\rho^{Ein}_0=0.42$ and $\rho^{Bur}_0=0.41$~GeV/cm$^3$.

As expected, the impact of the assumed DM density profile on the upper bounds is specially notable in the low mass region, where small changes in the tail of the speed distribution function can enhance or decrease prominently the number of expected DM events above the threshold~\cite{Strigari:2009zb,McCabe:2010zh,Frandsen:2011gi,Green:2011bv,Fairbairn:2012zs,Bernal:2014mmt}. Therefore, in the $m_{\chi}<30$~GeV region, our upper limits using well-motivated DM profiles and a self-consistent extraction of the speed distribution tend to be more stringent than using the SHM, as expected from figures~\ref{fig:rates_Ge} and~\ref{fig:rates_Xe}. These differences can be as high as one order of magnitude for masses below 10 GeV. Notice that the mean values of $v_{esc}$ we have used are in the three cases below 544~km/s, the value used for the SHM. This means that the difference in the upper limits cannot be ascribed to the different escape velocities assumed. In fact, this shows that the slope of the speed distribution at high velocities is smaller in the case of the NFW, Einasto and Burkert profiles. Needless to say that the values used for these distributions are subject to important uncertainties which actually in some cases would push these upper bounds closer to the SHM result.

\subsection{Limits on the SD cross section}
\label{sec:SD-m}

\begin{figure}[t!]
	\begin{center}  
	\epsfig{file=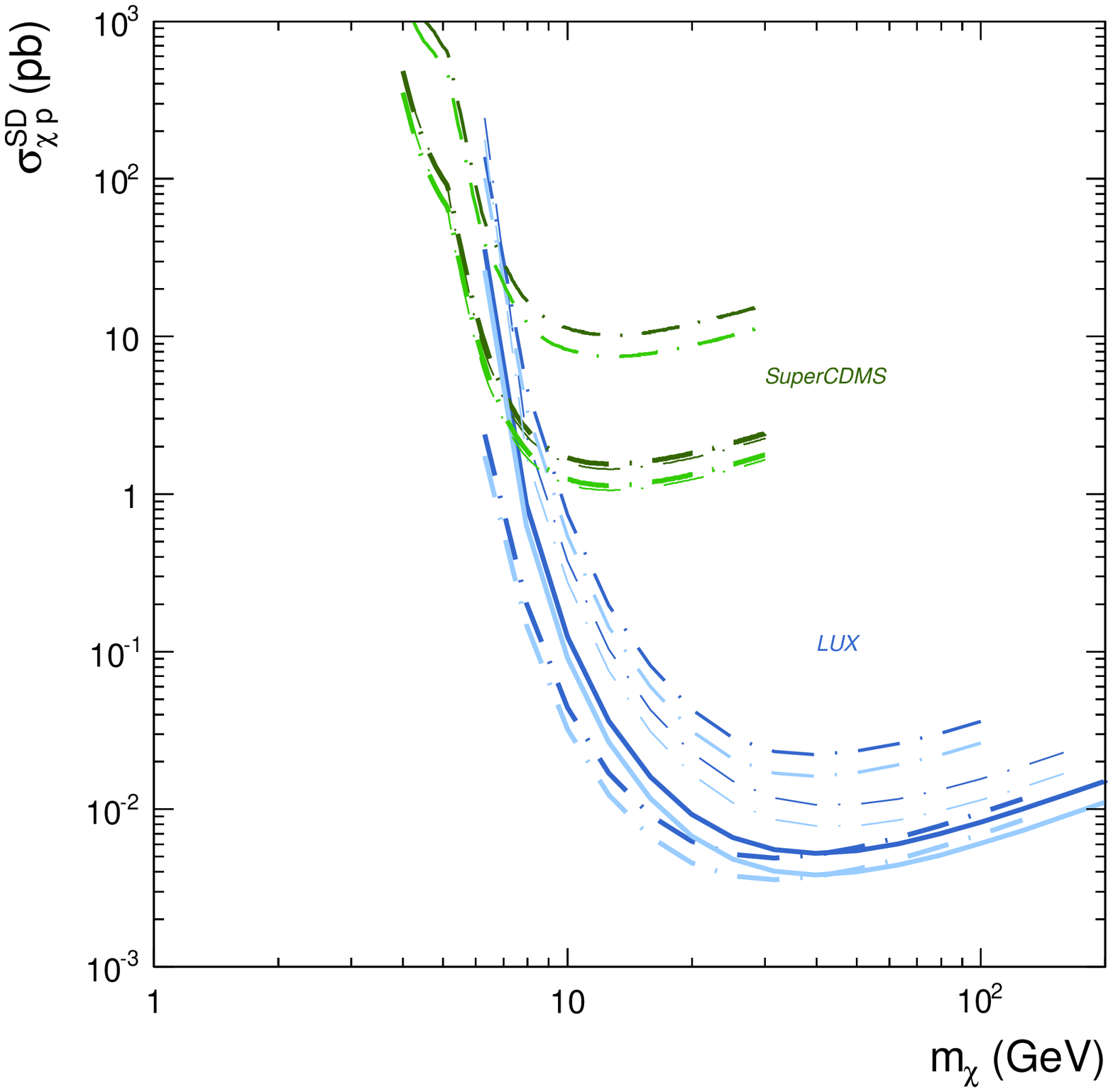,width=7.6cm}
	\epsfig{file=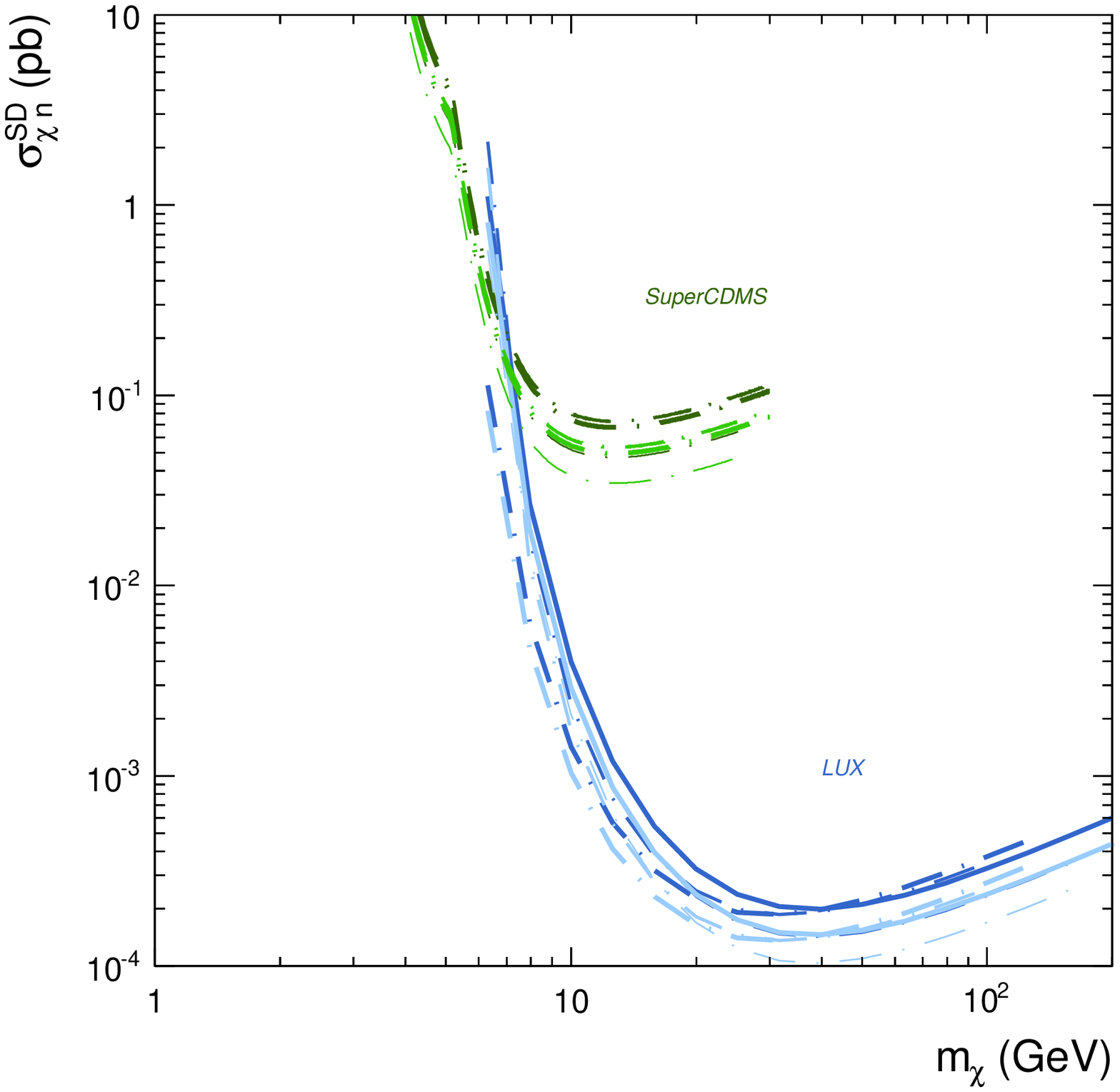,width=7.6cm}
	\end{center}
\caption{\small Upper bounds on the SD proton (right) and neutron (left) cross sections as a function of the DM mass for the LUX (blue curves) and SuperCDMS (green curves) experiments. The DM speed distributions used in each case are: SHM (solid), NFW (dashed-dotted). Darker colours correspond to $\rho_0=0.3$~GeV/cm$^3$ while lighter colours to the mean value of the local density extracted from Ref.~\cite{Fornasa:2013iaa}. The line width of each curve denotes the use of different nuclear structure functions. For LUX, the thick curve is chiral EFT, the medium is the Bonn A potential, and the thin is the Nijmegen potential. For SuperCDMS, the thick is the chiral EFT, the medium is the Dimitrov \textit{et al.} calculation and the thin is the Ressell \textit{et al.} calculation. See text for more details.
}
  \label{fig:sd-m}
\end{figure}

Now, we move to the case of SD interactions. Our results for the upper limits on the SD cross section of DM off protons (left panel) and neutrons (right panel) for the LUX and SuperCDMS experiments are displayed in figure~\ref{fig:sd-m}. We show the SHM (solid) and NFW (dashed-dotted curves) cases, where we have included different SD structure functions. Namely, we have employed the results obtained with chiral effective field theory (EFT) currents~\cite{Klos:2013rwa} for Ge and Xe isotopes (thicker lines), we have also used the Bonn A potential~\cite{Bednyakov:2006ux} for Xe and the Dimitrov \textit{et al.} calculation~\cite{Dimitrov:1994gc} for Ge (medium lines), and finally, the Nijmegen potential~\cite{Bednyakov:2006ux} for Xe and the Ressell \textit{et al.} calculation~\cite{Ressell:1993qm} for Ge (thinner lines). The importance of this kind of uncertainties on direct detection experimental results has been shown to be important in Ref.~\cite{Cerdeno:2012ix}, which motivates us to include the upper bounds for the different calculations of the SD structure functions in the literature. We have checked that these results are in good concordance with those of Ref.~\cite{Savage:2015xta} which were performed using a profile likelihood analysis.

We note from figure~\ref{fig:sd-m} that due to the nuclear spin of both Ge and Xe, the constraints from the neutron component of the cross section is substantially more stringent than that of protons. Nonetheless, the difference between distinct calculations of the SD structure functions is smaller in the neutron case. These two facts together imply that the exclusion of models from LUX and SuperCDMS results is not going to undergo strong variations from the choice of the SD structure functions. By contrast, from the left panel of figure~\ref{fig:sd-m} one can see that for both, LUX and SuperCDMS, the change in the $\sigma^{SD}_{\chi-p}$ owing to different SD structure functions can be of the order 10. Of course, this might be particularly important for models in which the DM particle couples to protons rather than to neutrons.  

\subsection{Combined limits for SI and SD interactions}

\begin{figure}[t!]
	\begin{center}  
	\epsfig{file=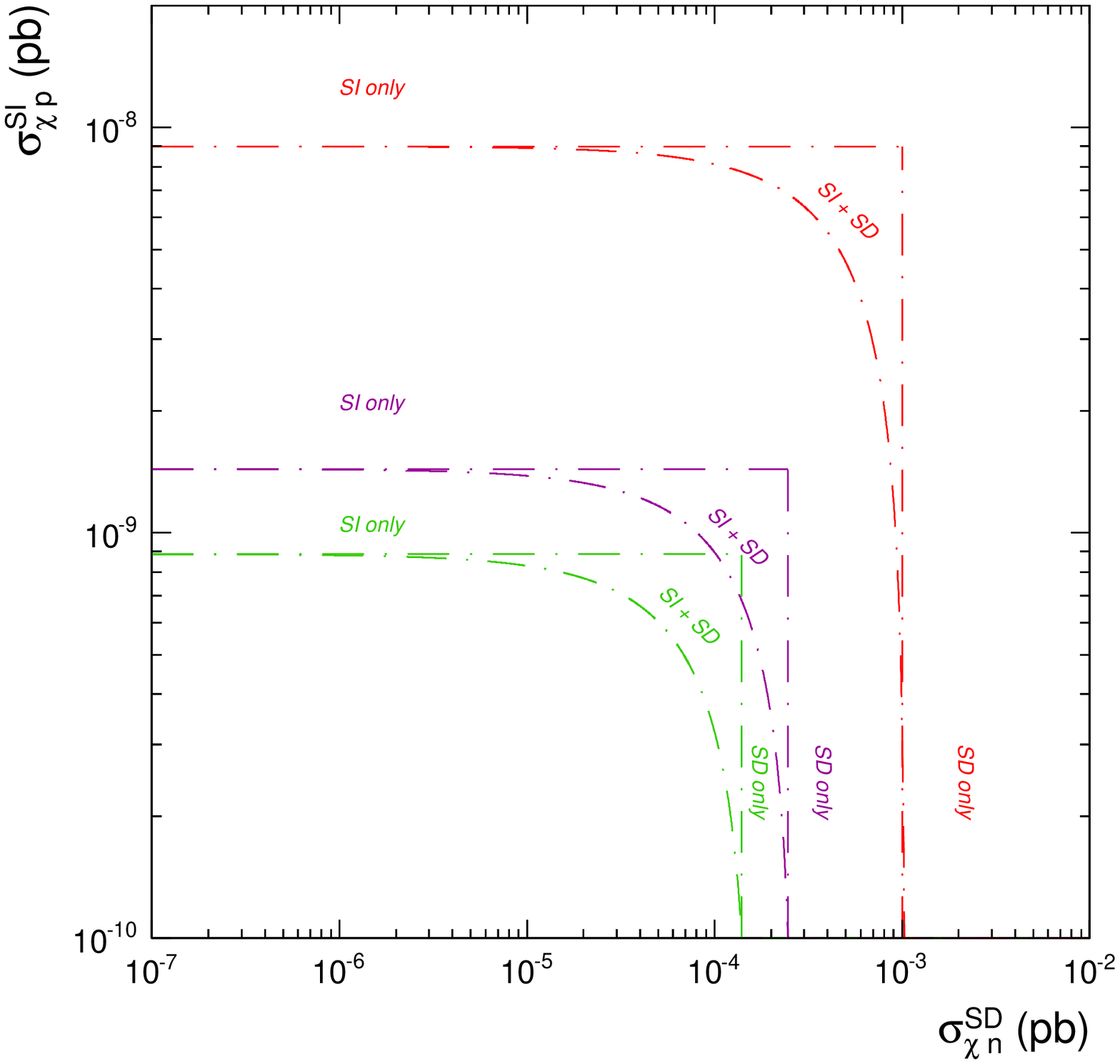,width=7.6cm}
	\epsfig{file=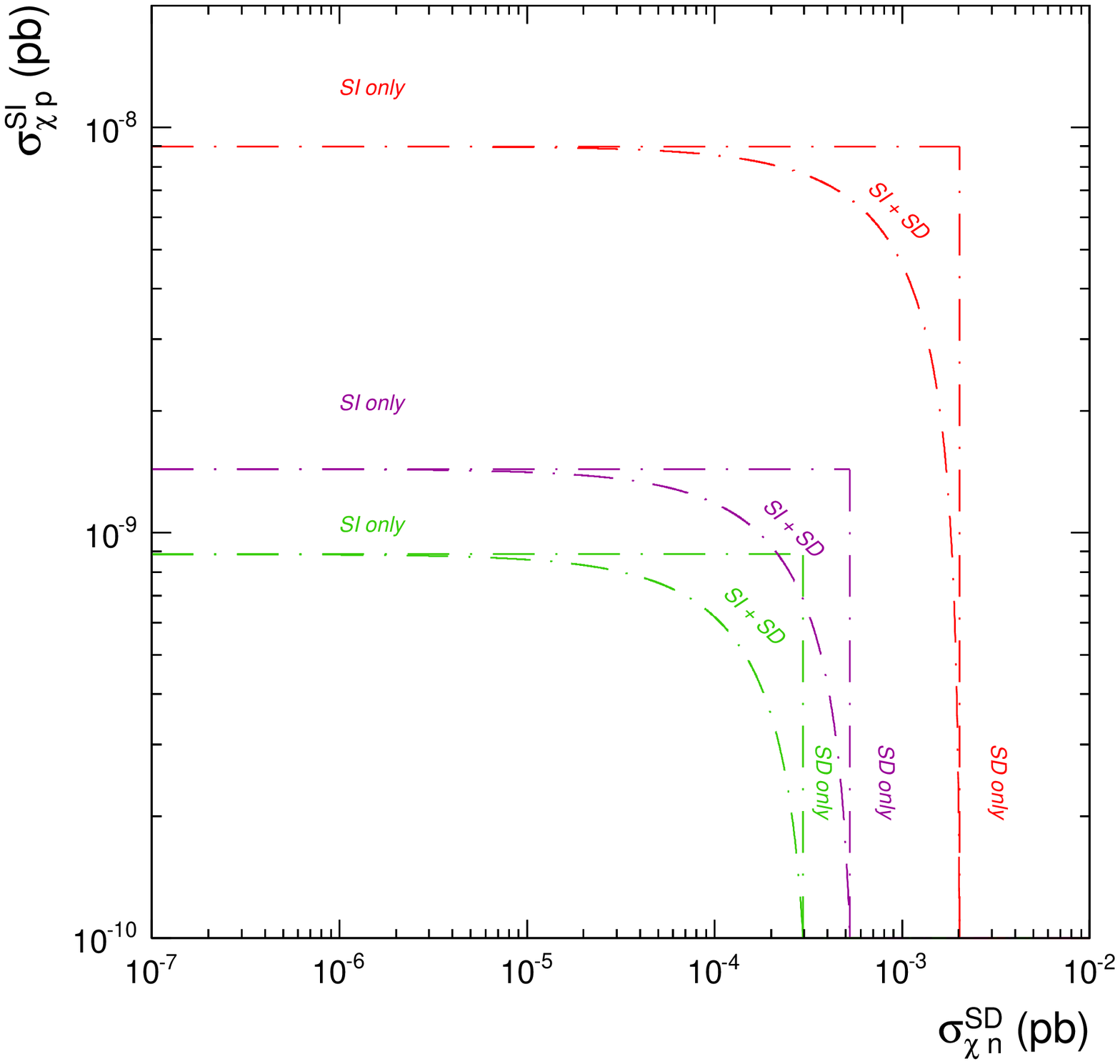,width=7.6cm}
	\end{center}
\caption{\small LUX combined SI and SD upper limits using a NFW profile for different masses, and $a_n/a_p=1$ (left) and $a_n/a_p=-1$ (right). The colour convention is as in Fig.~\ref{fig:rates_Xe}. Horizontal and vertical lines represent the limits for SI and SD interactions only, respectively. 
}
  \label{fig:si-sd}
\end{figure}

In this section, we derive the LUX limits for a specific choice of the DM mass, and then we calculate the upper bounds on $\sigma^{SI}_{\chi p}$ ,  $\sigma^{SD}_{\chi n}$ and the combined limits. We have assumed $f_n/f_p=1$ and $a_n/a_p=\pm1$.
We perform this exercise to evidence the difference of taking into account both cross sections at the same time instead of considering each one separately.

In Figure~\ref{fig:si-sd}, we show the LUX combined bounds for a NFW profile, using $\rho_0=0.3$~GeV/cm$^3$, and for the aforementioned neutron to proton ratios for the SI and SD interactions. The structure functions used in the SD case are those from the chiral EFT~\cite{Klos:2013rwa}. Besides, we have fixed the DM mass to: 10 GeV (red), 50 GeV (green) and 100 GeV (magenta curves). Horizontal lines are the SI limits considering zero contributions from the SD interactions, while vertical lines represent the opposite, i.e. SD limits with no contribution from the SI interactions. The rectangles delimited by these two kind of lines represent the cases in which the limits are considered for each interaction separately. 
Conversely, when one takes into account both interactions at the same time, each rectangle bends and a new excluded-to-be region enclosed by the rectangles and the new limits, labeled by SI $+$ SD, arises. Of course, a source of variation of these combined limits is given by the change of the ratios $f_n/f_p$ for SI interactions, and $a_n/a_p$ for the SD ones. For the latter, in Figure~\ref{fig:si-sd} we have plotted two different values,  $a_n/a_p=1$ in the left panel, and  $a_n/a_p=-1$ in the right panel. The difference among these two results highlights the importance that these quantities have in the upper bounds. Regarding the SI interactions, the effect can be also remarkable, specially when $f_n/f_p$ gets close to $-0.7$, i.e. the Xe-phobic region, but also for $|f_n/f_p|\gg1$.

Indeed, the aforesaid excluded regions are not taken into account if one applies the LUX results for the SI and SD interactions separately. Obviously, to extract information to rule out these regions it is necessary to compute the combined upper limits, but since the SI contribution is normally dominant, experimental collaborations only show results for one cross-section at a time. However, our results pose a question about the importance of the inclusion of such information to probe consistently DM models. As already stated, usually the SI contribution to direct detection experiments is dominant over that of the SD, because the former is summed coherently for all nucleons, hence scaling as $A^2$ (see Eq.~\ref{eq:zeromoment}). Nevertheless, there are scenarios in which the SD contribution can be of the same order than that of the SI, or even dominate over it~\cite{Bertone:2007xj}. In the following, we will show that proper combined limits are an important ingredient to take into account when analysing models. To such end, we will study two well motivated and popular DM models, the NMSSM and a $Z'$ portal.

\section{Next-to-Minimal Supersymmetric Standard Model}
\label{sec:nmssm}

The NMSSM is one of the most popular supersymmetric models (see for instance Ref.~\cite{Maniatis:2009re} for a review and Ref.~\cite{Belanger:2005kh} for a DM related analysis). In this model, the DM candidate is the lightest neutralino ($\tilde{\chi}^0_1$) which is, in general, a mixture of the superpartners of the gauge bosons and the Standard Model Higgs, plus a singlet scalar which is added in order to solve the $\mu$-problem. Since neutralinos are Majorana particles they contribute to both, SI and SD interactions, and thereby are perfect candidates to search for the aforementioned cases. 

Due to the complexity of the model, we will only focus on light neutralino masses, namely below $m_{\tilde{\chi}^0_1}<$~200~GeV. Furthermore, the speed distribution impact on the upper limits is maximised in this region.  Several analyses have addressed light neutralino DM in the NMSSM \cite{Gunion:2005rw,Vasquez:2010ru,Das:2010ww,Draper:2010ew,Cao:2011re,Carena:2011jy,AlbornozVasquez:2011js,Kozaczuk:2013spa,Huang:2013ima,Cheung:2014lqa,Huang:2014cla,Guo:2014gra,Bi:2015qva,Cao:2015loa}.

To look for phenomenologically viable solutions of this model, we have carried out a series of scans over the parameter space with {\tt MultiNest 3.9} \cite{Feroz:2007kg,Feroz:2008xx,Feroz:2013hea}. 
To this aim, we have defined a likelihood function whose 
inputs are the neutralino relic density, the SM Higgs mass, ${\rm BR}(B_s\to \mu^+\mu^-)$, and ${\rm BR}(b\to s\gamma)$, 
which are taken as gaussian probability distribution functions around the measured values with 
$2\sigma$ deviations. The gluino soft mass is fixed to $M_3 = 1500$~GeV, the trilinear parameters to $A_{U}=3700$~GeV, $A_{D}=2000$~GeV, and $A_{E}=-1000$~GeV, and the soft scalar masses of sleptons and squarks to 
$m_{\tilde{L}_i}=m_{\tilde{E}_i}=300$ GeV and
$m_{\tilde{Q}_i}=m_{\tilde{U}_i}=m_{\tilde{D}_i}=1500$~GeV, respectively, where
the index $i$ runs over the three families. This conservative choice of the squark masses is motivated by the LHC null searches. Also note that despite the high trilinear term $A_U$, the instability against charge- and/or color-breaking minima is avoided since the squark soft masses are at the TeV scale \cite{Ellwanger:1999bv}.

\begin{table}
  \begin{center}
    \begin{tabular}{|c|c|c|c|}

      \hline
      Parameter & Scan 1 & Scan 2&Scan 3\\
      \hline
      \hline 
      $M_1$& $1 - 200$ & $1 - 40$ & $1 - 200$\\
      $M_2$&  $200 - 1000$ & $200 - 1000$& $700 - 1000$\\
      $\tan\beta$ & $4 - 20$ & $4 - 20$ & $2 - 50$\\
      $\lambda$& $0.1 - 0.6$ & $0.1 - 0.6$& $0.001 - 0.1$\\
      $\kappa$& $0 - 0.1$ & $0 - 0.1$& $0.1 -0.6$\\
      $A_\lambda$& $500 - 5000$ & $500 - 5000$& $500 - 1100$\\    
      $A_\kappa$& $-50 - 50$ & $-30 - 0$& $-50 - 50$ \\    
      $\mu_{eff}$& $110 - 250$ & $160 - 250$& $200 - 400$\\   
      \hline
      $M_3$ & $1500$ & $1500$&$1500$\\ 
      $A_{U}$ & $3700$ & $3700$&$3700$\\ 
      $A_{D}$ & $2000$ & $2000$&$2000$\\ 
      $A_{E}$ & $-1000$ & $-1000$&$-1000$\\
      $m_{\tilde{L}_i}=m_{\tilde{E}_i}$ & $300$ & $300$& $300$\\ 
      $m_{\tilde{Q}_i}=m_{\tilde{U}_i}=m_{\tilde{D}_i}$ & $1500$ & $1500$& $1500$\\
      \hline
    \end{tabular}
    \caption{Input parameters for the series
      of NMSSM scans. Masses and trilinear parameters are given in GeV. All parameters are defined at the EW scale.}
    \label{tab:scan_nmssm}
  \end{center}
\end{table} 

We have performed three scans over the
parameter space where the different input parameters are varied
according to Table\,\ref{tab:scan_nmssm}. The neutralino relic abundance has been calculated with {\tt micrOMEGAs 3.6.9}~\cite{Belanger:2013oya}.
We have used {\tt NMSSMTools 4.1.2}~\cite{Ellwanger:2004xm,Ellwanger:2005dv,Ellwanger:2006rn} to compute the NMSSM mass spectrum, the masses of the Higgs bosons including full two-loop contributions, and the relevant 
low-energy phenomenology observables. LHC measurements of the Higgs properties are included by constructing the $\Delta\chi^2$ distribution from Ref.~\cite{Belanger:2013xza} and allowing $3\sigma$ deviations. Limits on the velocity averaged cross section of DM particles from the gamma-ray null observation of dwarf spheroidal galaxies (dSphs) are also taken into account, following the procedure sketched in Ref.~\cite{Cerdeno:2015ega} for the Pass 8 data~\cite{Ackermann:2015zua}. In addition, we have implemented the recent measurements of the branching ratio of the $B_s\to \mu^+\mu^-$ process
by the LHCb \cite{Aaij:2013aka} and CMS \cite{Chatrchyan:2013bka} Collaborations, which collectively yield
$1.5\times 10^{-9}< {\rm BR}(B_s\to \mu^+\mu^-)< 4.3\times 10^{-9}$ at
95\% CL. 
For the $b\to s\gamma$ decay, we have considered the 2$\sigma$ range
$2.89\times 10^{-4}< {\rm BR}(b\to s\gamma)< 4.21\times
10^{-4}$, which takes into account
theoretical and experimental uncertainties added in quadrature \cite{Ciuchini:1998xy,D'Ambrosio:2002ex,Amhis:2012bh,Misiak:2015xwa,Czakon:2015exa}.
We have also imposed $0.85\times 10^{-4}< {\rm BR}(B^+ \to \tau^+ \nu_\tau)<2.89\times 10^{-4}$ \cite{Lees:2012ju}.

Finally, we have set an upper bound on the neutralino relic abundance, $\Omega_{\tilde{\chi}^0_1} h^2<0.13$,
consistent with the latest Planck results~\cite{Ade:2013zuv}. 
Besides, we have considered the possibility that neutralinos might only contribute to a fraction of the total relic density. To deal with these cases, the fractional density, $\xi=\min[1,\Omega_{\tilde{\chi}^0_1} h^2/0.11]$, have been introduced to account for the reduction in the rates for direct and indirect searches\footnote{Although the introduction of $\xi$ for certain cases of multicomponent DM, such as self-interacting DM, might not be consistent.}.  

\subsection{Combined limits for different speed distributions}

\begin{figure}[t!]
	\begin{center}  
	\epsfig{file=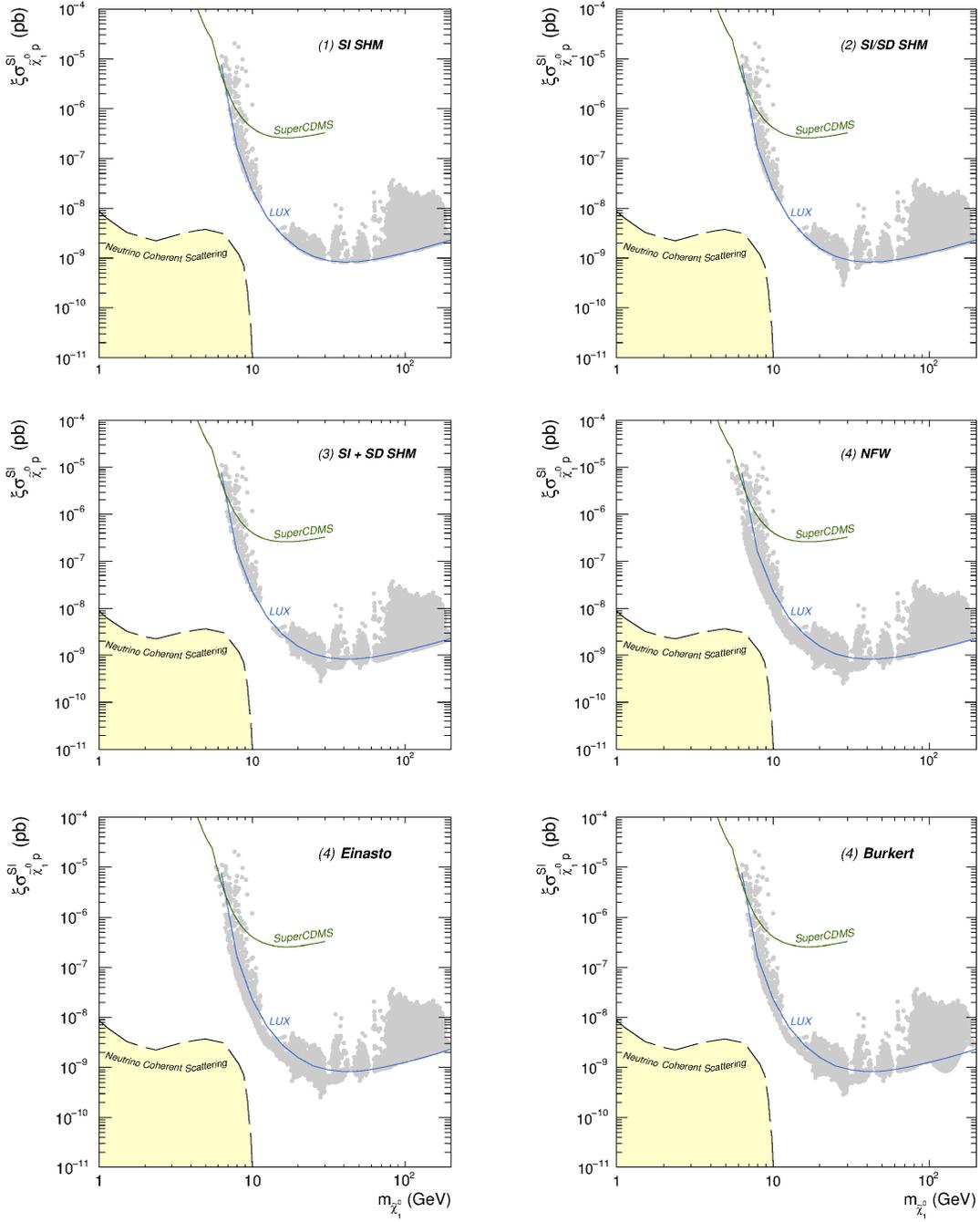,width=14.6cm}
	\end{center}
\caption{\small Excluded points for neutralino DM in the NMSSM in the plane $\xi\sigma^{SI}_{\tilde{\chi}^0_1-p}$-$m_{\tilde{\chi}^0_1}$ using different assumptions on the LUX and SuperCDMS limits and different speed distributions. The blue (green) curve represents the LUX (SuperCDMS) bound only for SI interactions and for the SHM.  
}
  \label{fig:si-NMSSM-halos}
\end{figure}

In this section, we will discuss how the inclusion of direct detection upper limits taking into account the SI and SD interactions at the same time can have an impact on the viability of DM scenarios in the NMSSM that otherwise will be allowed. More precisely, we will use the results from previous sections to illustrate how different ways of including direct detection bounds can affect the parameter space of the NMSSM.  We will  also show the impact of the astrophysical uncertainties from the speed distributions, leaving to the next section the  analysis of the effect of these uncertainties on the SD structure functions. 

To show schematically how different solutions are excluded depending on the assumptions considered when evaluating the limits, we will use the following procedure:\\
\noindent (1) \textbf{SI}: We will apply first the SuperCDMS and LUX bounds only for the SI interactions and $f_/f_p=1$ to the viable solutions found, using the SHM. This bound correspond to the usual bound published by the SuperCDMS and LUX Collaborations.\\
\noindent (2) \textbf{SI/SD}: Then, we will add to the previous limits the bounds that only take into account the SD interactions, also for the SHM. This step is equivalent to apply the combined limits generating the rectangles shown in Fig.~\ref{fig:si-sd}.\\
\noindent (3) \textbf{SI $+$ SD}: Next, instead of the previous considerations we will calculate the bounds taking into account all the contributions, SI and SD and the specific neutron to proton ratios. This is also done for the SHM.\\
\noindent (4) \textbf{Profile}: Same as (3) but for the NFW, Einasto and Burkert velocity distributions.\\

\begin{figure}[t!]
	\begin{center}  
	\epsfig{file=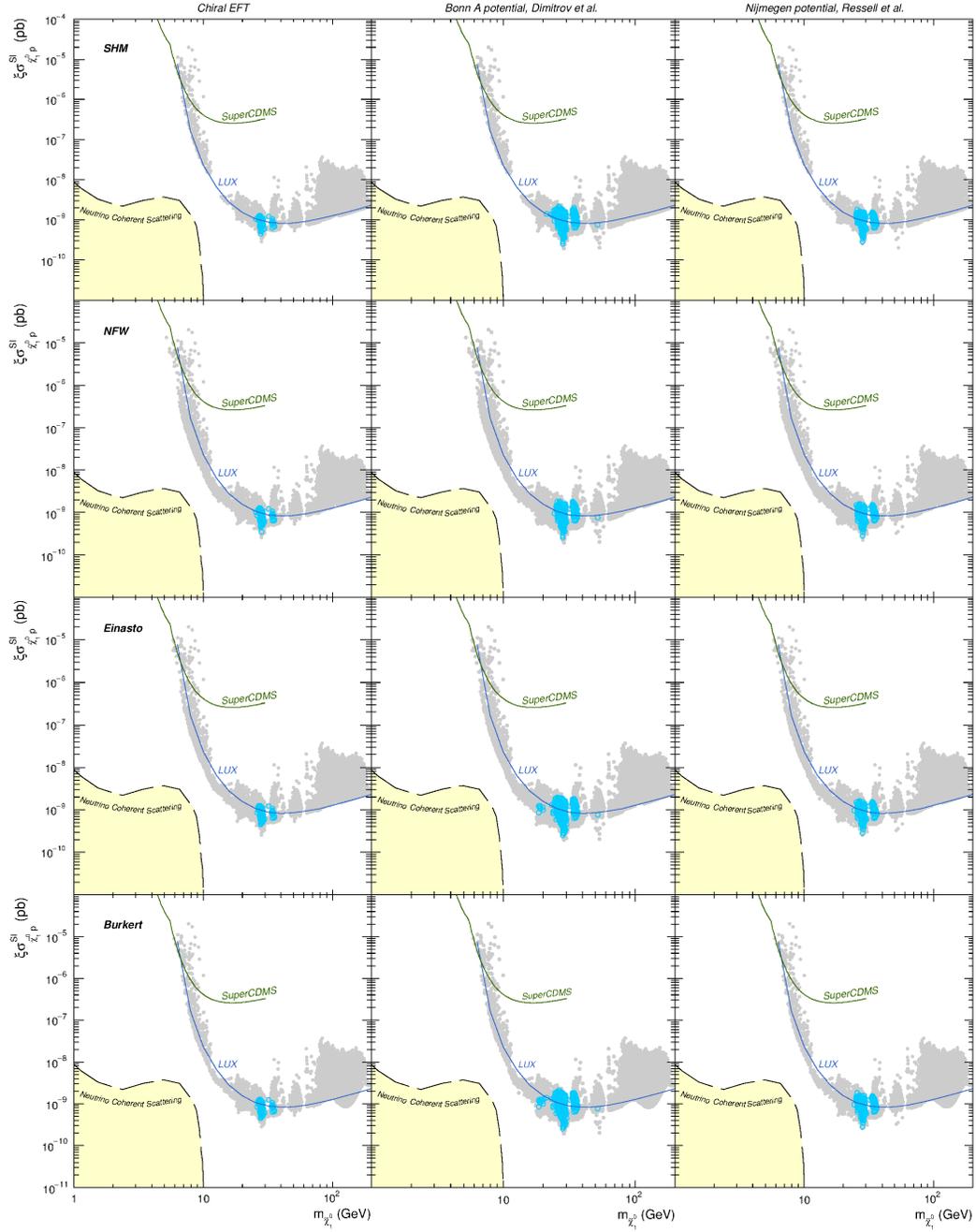,width=14.0cm}
	\end{center}
\caption{\small Excluded points for neutralino DM in the NMSSM in the plane $\xi\sigma^{SI}_{\tilde{\chi}^0_1-p}$-$m_{\tilde{\chi}^0_1}$ using different speed distributions and SD structure functions. 
The blue (green) curve represents the LUX (SuperCDMS) bound only for SI interactions and for the SHM. 
Light blue points denote solutions in which the contribution to the expected number of events in LUX is dominated by SD interactions.
}
  \label{fig:si-NMSSM-ff}
\end{figure}

Notice that 
with the purpose of isolating the effect that speed distributions have on the exclusion regions, 
in none of the previous steps we have varied the value of the local density, $\rho_0$. 
%The reason for this is to isolate the effect that the speed distribution has on the exclusion regions. 
In fact, if we change the local density value in agreement with those extracted in Ref.~\cite{Fornasa:2013iaa}, according to Fig.~\ref{fig:si-m}, the upper bounds become more stringent.

The set of plots shown in Figure~\ref{fig:si-NMSSM-halos} depicts how the previous steps affect the exclusion regions of neutralinos in the NMSSM. All the plots show the points excluded by the SuperCDMS and  LUX results in the plane $\xi\sigma^{SI}_{\tilde{\chi}^0_1-p}$-$m_{\tilde{\chi}^0_1}$. In the first row left-hand side plot, we display the solutions excluded when the SI proton cross section weighted by $\xi$ is above the nominal SuperCDMS and LUX limits. This corresponds to step (1), as labelled in the plot. In the same row, the right-hand side plot, includes also the upper bounds both for SD proton and neutron interactions. As it can be seen, when compared with respect to the left-hand side plot, a new population of excluded points has appeared around $m_{\tilde{\chi}^0_1}\approx 30$~GeV. These points correspond to solutions in which the SD neutron cross section is above the LUX bound\footnote{Remind that Xe-based experiments are much more sensitive to the neutron contribution of the SD interactions since the total spin of these nuclei are  dominated by neutrons.}.  

Next, we move to the middle row of Figure~\ref{fig:si-NMSSM-halos}. In the left panel, we note that many solutions previously allowed by LUX are now ruled out. These points fall precisely in the regions labelled by SI + SD in Fig.~\ref{fig:si-sd}. All the points that have appeared at this step have a similar contribution from the SI and SD interactions to the expected number of events in LUX. 
Therefore, all these solutions require a careful calculation of the LUX upper bound to be correctly excluded. This nicely highlight the importance of using combined limits and showcases how the upper limits should be compared point per point in the parameter space. This task can be achieved very quickly with the help of the tables provided in this work.

The remaining plots correspond to step (4). 
As previously discussed, the impact of considering different velocity distributions is notable for neutralino masses below approximately 50 GeV, as expected from the results shown in Fig.~\ref{fig:si-m}. 
This light mass region is actually one of the favoured regions that explain the Fermi-GeV excess in terms of annihilating DM particles~\cite{Vitale:2009hr,Hooper:2010mq,Morselli:2010ty,Hooper:2011,Abazajian:2012pn,Daylan:2014rsa,Gordon:2013vta,Abazajian:2014fta,Zhou:2014lva,Calore:2014xka}. 
Moreover, this excess prefers a NFW-like profile; hence, the use of realistic speed distributions according to the Galactic DM profile is required to suitably constraint possible DM explanations of this excess 
 with direct detection upper bounds 
%can lead to important changes in direct detection upper bounds
~\cite{Peiro}. 

Thus, our findings point out that in order to explore robustly the parameter space of the NMSSM, one has to include properly the SuperCDMS and LUX limits.
It is worth noting that steps (1) and (2) assume a fixed neutron to proton ratio for both interactions, which does not coincide exactly with those of neutralinos, when we apply step (3) this subtlety is considered using the exact values that neutralinos have for these quantities. 
Therefore, these ratios can affect the exclusion of some of the previously allowed points. 

\subsection{Combined limits for different SD structure functions}

As seen before, many solutions found in our scans predict a SD contribution similar to that of the SI. 
In Section~\ref{sec:SD-m}, we have shown that the use of distinct structure functions lead to differences in the LUX upper bounds that can be significant. 
Therefore, we must expect that the choice of the structure functions is going to impact on the excluded regions of neutralinos in the NMSSM.   

In Figure~\ref{fig:si-NMSSM-ff}, we depict the excluded points assuming different SD structure functions (columns) and speed distributions (rows). 
For the sake of completeness,  we have plotted in light blue those points in which the contribution to the DM expected events in LUX from the SD interactions is dominant. As we can observe, in Figure~\ref{fig:si-NMSSM-ff} from left to right panels, light blue points are generally affected by the choice of the structure functions and, in agreement with Figure~\ref{fig:sd-m}, the chiral EFT structure function is able to rule out smaller regions of the parameter space. It is noteworthy that the BonnA potential model of the SD structure functions provides the most stringent limits of Xe-based experiments, highlighted as a higher population of light blue points in the middle column of Figure~\ref{fig:si-NMSSM-ff} with respect to the other two (compare also steps (2) and (3) in Figure~\ref{fig:sd-m}). Notice that the SD dominant solutions we have found accumulate around neutralino masses of 30 GeV, while those with similar SI and SD contributions in LUX range from 20 to 100 GeV approximately (see also in Fig.~\ref{fig:si-NMSSM-halos} the change from step (2) to step (3)). This means that the choice of the SD structure function has a notable effect on the allowed (excluded) solutions for NMSSM neutralinos below 100 GeV, which is precisely the region where most of the explanations of the Fermi-GeV excess in this model lie~\cite{Cheung:2014lqa,Guo:2014gra,Bi:2015qva,Cao:2015loa,Butter:2015fqa}.

\section{$Z'$ portal}
\label{sec:zprime}

We now turn our attention to another popular and well-motivated extension of the SM, the $Z'$ portal. In this scenario, a hidden sector communicates with the SM via a gauge boson, provided that the SM is enlarged with an extra abelian gauge group~\cite{Langacker:2008yv}. The phenomenology of these constructions is very rich, and ranges from colliders to direct and indirect searches for DM~\cite{Dudas:2009uq,Cassel:2009pu,Frandsen:2011cg,Barger:2012ey,Arcadi:2013qia,Alves:2013tqa,Foot:2014uba,Foot:2014osa,Cline:2014dwa,Alves:2015pea,Martin-Lozano:2015vva,Chala:2015ama,Alves:2015mua}.

We will consider the same construction as in Ref.~\cite{Martin-Lozano:2015vva}, which is based on a certain type II string compactifications with intersecting branes. The key point of these scenarios is that  gauge bosons acquire a mass through the St\"uckelberg mechanism~\cite{Feng:2014eja,Feng:2014cla}, except of course the one corresponding to the hypercharge, which remains massless in the phase of unbroken electroweak symmetry.

\begin{table}
  \begin{center}
    \begin{tabular}{|c|c|c|c|}

      \hline
      Parameter & Range\\
      \hline
      \hline 
      $|a|$& $10^{-4} - 1$ \\
	  $|b|$& $10^{-4} - 1$ \\
	  $|c|$& $10^{-4} - 1$ \\
	  $|d|$& $10^{-4} - 1$ \\
	  $|h|$& $10^{-4} - 1$ \\
	  $m_{\psi}$& $1 - 200$ \\
	  $m_{Z'}$& $500 - 8000$ \\	  	  	  
      \hline
    \end{tabular}
    \caption{Input parameters for the scan of the $Z'$ portal model. Masses are given in GeV.}
    \label{tab:scan_z}
  \end{center}
\end{table} 

We will assume throughout this section that the DM in this scenario is a Dirac fermion, $\psi$, which lives in the hidden sector. The phenomenology of this kind of DM will be driven by the interaction with the SM fermions through the exchange of a $Z'$ boson. The couplings of the $Z'$ to the SM fermions will be determined by specific combinations of four parameters, $a$, $b$, $c$ and $d$~\cite{Martin-Lozano:2015vva}.  We have performed a scan over the parameter space, with the parameter ranges given in Table~\ref{tab:scan_z}, where $h$ is the coupling of DM to the $Z'$ boson. To this end, we have implemented the model in {\tt micrOMEGAs 3.6.9}~\cite{Belanger:2013oya}. We have taken into account that there exist certain combinations of the these parameters that are not allowed by the orthogonality and normalization conditions imposed on the $Z'$ mass eigenvectors~\cite{Martin-Lozano:2015vva}. 

\begin{figure}[t!]
	\begin{center}  
	\epsfig{file=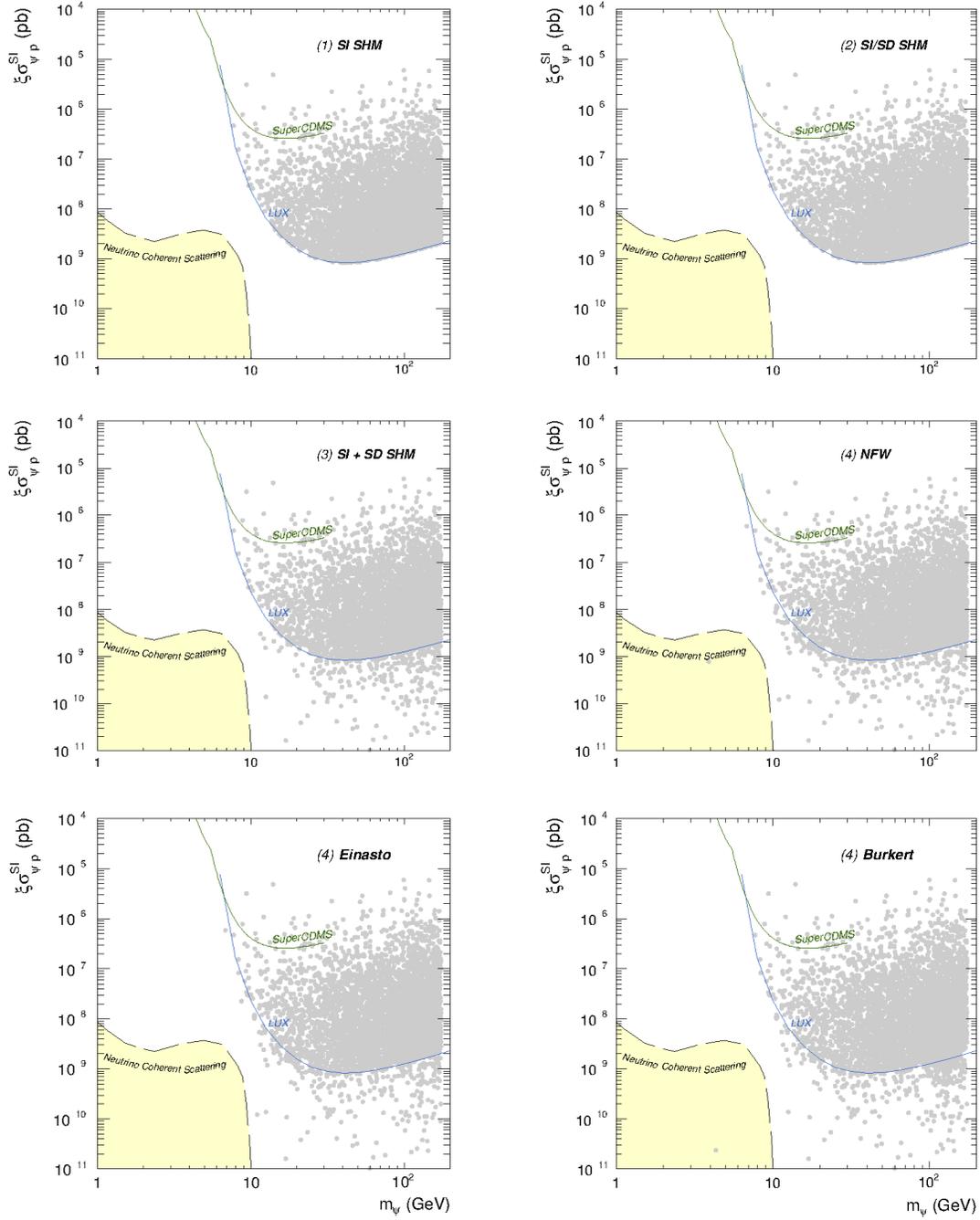,width=14.6cm}
	\end{center}
\caption{\small Excluded points for Dirac DM in the $Z'$ portal model in the plane $\xi\sigma^{SI}_{\psi-p}$-$m_{\psi}$ using different assumptions on the LUX and SuperCDMS bounds and different speed distributions. 
The blue (green) curve represents the LUX (SuperCDMS) bound only for SI interactions and for the SHM.  
}
  \label{fig:si-zp-halos}
\end{figure}

In general, $U(1)$ extensions of the SM can be constrained using resonances at colliders, since the coupling of the $Z'$ boson to leptons and quarks contributes to the appearance of dimuon, dielectron and dijet resonances. Here, we have applied the ATLAS results for high mass resonances decaying into a $\mu^+\mu^-$ or an $e^+e^-$ pair at a center of mass energy $\sqrt{s}=8$~TeV and luminosities of $20.5$~fb$^{-1}$ and $20.3$~fb$^{-1}$ for dimuons and dielectrons resonances, respectively~\cite{Aad:2014cka}. Furthermore, we have used the searches for dijet resonances and monojets plus missing energy that receive additional contributions from the presence of a $Z'$ boson, both at the LHC and Tevatron colliders~\cite{Aaltonen:2008dn,Aad:2011fq,CMS-dijet}. In order to implement these constraints, we have followed the procedure of Ref.~\cite{Martin-Lozano:2015vva}. 

It is worth mentioning that we do not impose any bound on the relic abundance of $\psi$. This is a consequence of the potential complexity of the gauge structure and matter content of the hidden sector, which might induce other contributions that account for the correct abundance such annihilations of $\psi$ into hidden sector matter. Of course, this will affect the predictions of the model for indirect dark matter searches. To deal with this, we have defined $\xi$ as in the NMSSM, but in this case if $\Omega_{\psi}h^2>0.13$ then we set $\xi=1$.\footnote{This is equivalent to consider that whatever hidden processes there could be behind the relic density, $\psi$ particles can account for the 100\% of the DM in the Universe. } 

Finally, we have imposed the upper bounds on the annihilation cross section of $\psi$ from dSph galaxies using the Pass 8 data of the Fermi-LAT satellite~\cite{Ackermann:2015zua}. These limits are presented for specific SM final states, while in this model each point of the parameter space entails DM annihilation into a combination of different SM particles. This circumstance prevents us to use these Fermi-LAT results directly, but we have to weight each upper limit by the corresponding annihilation percentage instead.

\subsection{Combined limits for different speed distributions}

As in the NMSSM case, presented in Section~\ref{sec:nmssm}, in this section we will evaluate the impact of the speed distribution on the experimentally allowed solutions of the $Z'$ portal with Dirac DM. We will follow the same steps described before for the NMSSM.

In figure~\ref{fig:si-zp-halos}, we show the results for the $Z'$ portal model using the same steps as in figure~\ref{fig:si-NMSSM-halos}. Notably, from step (2) to (3) many new excluded solutions appear below the LUX SHM limit (blue curve). Unlike in the NMSSM, all these ruled out points correspond to solutions in which the ratio $f_n/f_p$ is different from 1, and in many of these cases, the neutron contribution to the SI cross section is much higher than that of the proton. For this reason, many points that lie far below the LUX bound on the SI-proton cross section are excluded. In addition, some of the solutions that were excluded in step (2) are now allowed. These solutions correspond to those with $f_n/f_p\approx-0.7$, the region in which the sensitivity of Xe-based experiments substantially decreases, the so-called Xe-phobic region. Also, solutions with $|f_n/f_p|<1$ are now allowed, since the neutron contribution does not reach the size of the proton one, and hence, the upper bound weakens.

Regarding the impact of the velocity distribution, one can see in figure~\ref{fig:si-zp-halos} that from step (3) on, more points are excluded in the low mass region. Although in this case, the difference is less visible than in the NMSSM, because the density of points for masses below 30 GeV is not very high. Furthermore, the fact that in this model $f_n/f_p$ can virtually take any value, makes more difficult to identify points where the effect of the speed distribution is remarkable. 

\subsection{Combined limits for different SD structure functions}

\begin{figure}[t!]
	\begin{center}  
	\epsfig{file=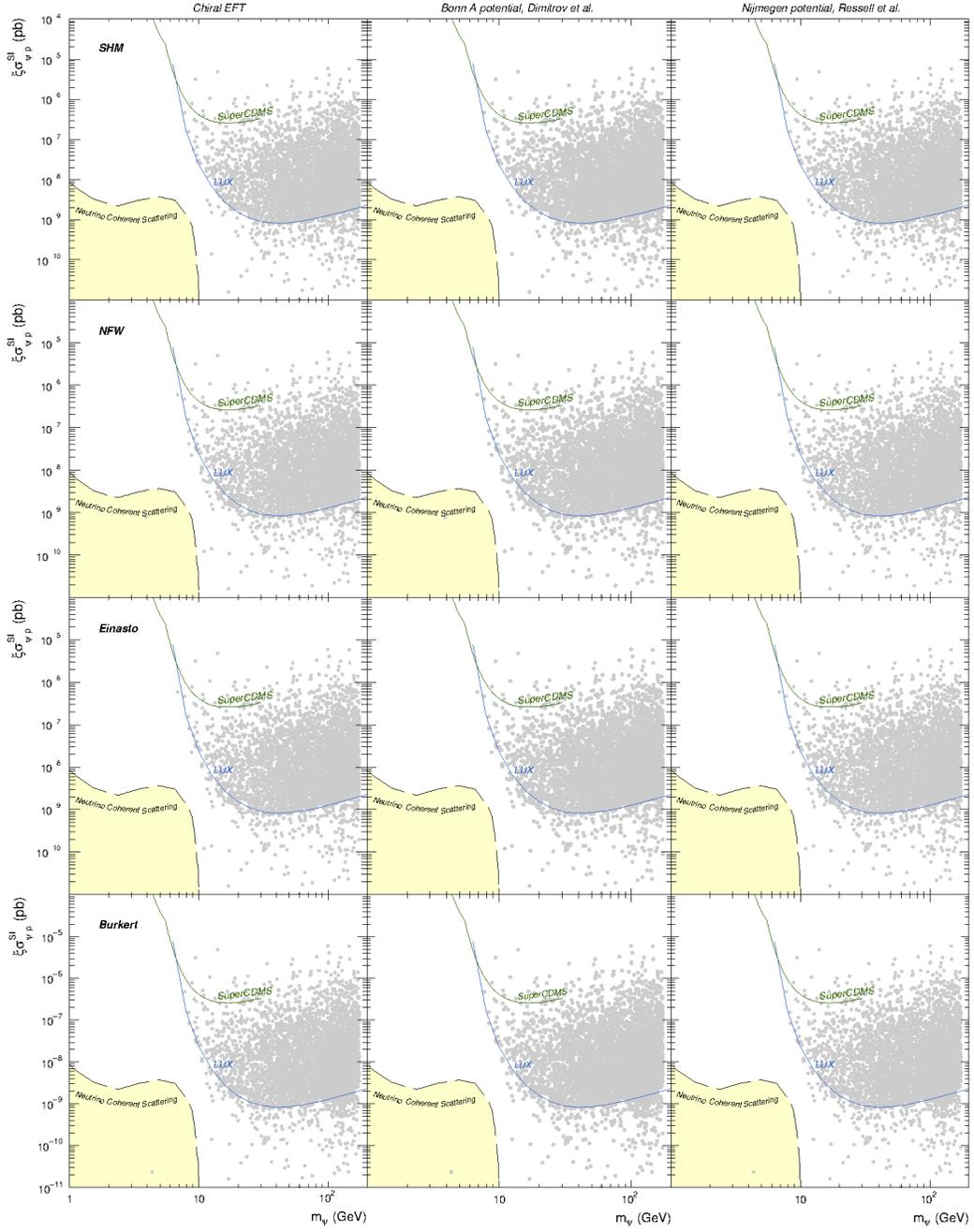,width=14.0cm}
	\end{center}
\caption{\small Excluded points for Dirac DM in the $Z'$ portal model in the plane $\xi\sigma^{SI}_{\psi-p}$-$m_{\psi}$ using different speed distributions and SD structure functions. 
The blue (green) curve represents the LUX (SuperCDMS) bound only for SI interactions and for the SHM. 
Light blue points denote solutions in which the contribution to the expected number of events in LUX is dominated by SD interactions.
}
  \label{fig:si-zp-ff}
\end{figure}

In figure~\ref{fig:si-zp-ff}, we show the results for the $Z'$ portal model using the same speed distributions and SD structure functions as in figure~\ref{fig:si-NMSSM-ff}. Notice that the absence of light blue points (like those appearing in figure~\ref{fig:si-NMSSM-ff} for neutralino DM) means that in this case we have not found any point of the parameter space in which the SD contribution to the total number of expected events in LUX dominates. Consequently, we observe that the change in the SD structure functions employed to impose the limits does not affect the exclusion of the solutions found.

\section{Conclusions}
\label{sec:conclusions}

In this paper, we have performed a thorough study of the impact of the assumptions usually made to compute DM direct detection upper bounds. We have used the NMSSM and a $Z'$ portal DM models to illustrate how the use of direct detection limits, including SI and SD interactions at the same time, different neutron to proton ratios, distinct DM density profiles as well as different SD structure functions, has important implications on the excluded parameter spaces. 

We have extracted the LUX and SuperCDMS upper bounds under many different assumptions about the speed distribution and SD structure functions. For the former, we have used the velocity distributions derived in Ref.~\cite{Fornasa:2013iaa} for three different choices of the DM density profile. In Ref.~\cite{Fornasa:2013iaa}, the authors computed the speed distribution in a self-consistent way from the gravitational potential of the Milky Way. We have employed these distributions, to compute upper limits that rely on a realistic description of the DM halo, and hence, produce results consistent with a wide range of astronomical observations. As expected, the differences in the upper bounds from the speed distribution focus mainly on the light mass range, $m_{DM}\lesssim50$~GeV. Another source of uncertainty for direct detection experiments are the SD structure functions. By using the different results in the literature, we have explicitly shown the variation in the upper limits due to the choice of the SD structure functions. 

In order to incorporate all these ingredients when analysing models, in each of the mentioned cases, we have carried out a decomposition of the expected DM signal in both experiments, LUX and SuperCDMS, in terms of several expressions that can be used to derive upper limits using statistical tools based on the total number of expected events. In this paper, we have used the maximum gap method, and obtained tabulated data, that is suitable for this method, for each of the different combinations of these ingredients. This data set allows to constrain any generic model  with both SI and SD interactions, for protons and neutrons. The data and python scripts that exemplify their use can be downloaded from this site \url{http://goo.gl/1CDFYi}.

As case studies to apply our findings, we have chosen the NMSSM and the $Z'$ portal with Dirac DM. We have performed a series of scans over the parameter space of both models imposing the latest experimental limits. More specifically, we have applied to the resulting points the constraints from the LHC and Tevatron, Planck, low energy observables and Fermi-LAT bounds from dSph galaxies. Then, we have imposed how the inclusion of the upper limits from LUX and SuperCDMS taking into account the SI and SD interactions, and different assumptions about the halo profile and SD structure functions, to analyse the effect of each of these factors on the allowed/excluded regions of the parameter space. Our findings point out that a more careful implementation of direct detection limits is needed when analysing models in light of the current experimental constraints. More specifically, in the NMSSM, the nature of the solutions found requires a combined implementation of the SI and SD bounds at the same time, since many solutions have similar contributions from both types of interaction. For this reason, in this model, not only the halo profile choice, but also the SD structure functions used to derive the SD contribution play an important role. In the case of the $Z'$ portal, the necessity of applying combined limits arises from the neutron and proton contributions to the SI interactions. This model, with a Dirac DM candidate, shows strong deviations from the $f_n=f_p$ equality. This situation is translated into huge variations of direct detection upper limits that should be considered for a careful analysis of the parameter space. 

The amount of experimental data nowadays is enormous. Huge experimental and theoretical efforts are being made to interpret these data in terms of many different models. With this purpose, theoreticians are performing very fine calculations of observables that can be constrained in light of the current data. In this sense, our results provide a missing piece to this effort, making easier the implementation of direct DM detection limits using more realistic assumptions; namely the SI and SD contributions at the same time, speed distributions in agreement with astrophysical observations, and different SD structure functions.  

\noindent{\bf \large Acknowledgements}

We thank D. G. Cerde\~no, M. Fornasa and F. Kahlhoefer for useful comments on the manuscript.
S.R. acknowledges the support of the Campus of Excellence UAM+CSIC.
M.P. thanks the support of the Consolider-Ingenio 2010 program under grant MULTIDARK CSD2009-00064 and the European Union under the ERC Advanced Grant SPLE under contract ERC-2012-ADG-20120216-320421. We thank the support of the Spanish MICINN under Grants No. FPA2012-34694 and FPA2013-44773-P, and the Spanish MINECO Centro de Excelencia Severo Ochoa Program under Grant No. SEV-2012-0249.

\appendix

\section{Extraction of upper limits using the maximum gap method}
\label{sec:functions}

In this appendix, we give detailed information about the extraction of the direct detection bounds for SuperCDMS and LUX using the data provided in this work. 
Notice that this data set can be used with two different, nevertheless equivalent, purposes. 
Its first application is to calculate the upper limits under a number of assumptions, 
in the same way the collaborations present their results for the SI proton cross section as a function of the mass (see for instance Fig.~\ref{fig:si-m}). 
The second application is to test if a given point in the parameter space of a model is allowed or excluded, as we have done to obtain Figs~\ref{fig:si-NMSSM-halos}-\ref{fig:si-zp-ff}. 

As stated in the Introduction, by means of these tabulated data one can determine the expected number of DM events in SuperCDMS and LUX in an easy and quick way. 
Then, with these number of events, we can estimate the upper bounds on DM interactions with protons and/or neutrons, using the Yellin's maximum gap method~\cite{Yellin:2002xd}, 
which produces excellent results (see Fig.~\ref{fig:si-m-comparison} for a comparison between our results and those of the experimental collaborations).
With the attached data,  one can calculate the expected number of events in specific energy windows and the total number expected for a given experiment. 
The former and the latter are necessary to apply this method to SuperCDMS 
(for LUX, we assume zero observed candidate events and then we will only need the total number of expected events).

Let us start considering a generic direct detection experiment, denoted by $Z$. It is straightforward to show that the number of expected DM events for the SI interactions can be written as, 
 \begin{equation}
N_{SI}=\sigma^{SI}_p\left[ F^{p}_{Z}(m_{\chi})+\left(\frac{f_n}{f_p}\right)^2 F^{n}_{Z}(m_{\chi})+\frac{f_n}{f_p}F^{pn}_{Z}(m_{\chi})\right],
\label{eq:nsi}
\end{equation}
where the $F^{p,n,pn}_{Z}$ functions enclose all the information about the DM halo, form factors, energy resolution, as well as the transformation from the actual measured energy to recoil energy (see Section~\ref{sec:basics}). 
These SI functions have been normalized to a reference cross section of $10^{-8}$~pb. 
Note that these functions have been calculated for a specific energy window, and then, are not valid for arbitrary energy intervals.
Furthermore, we have reabsorbed all the constants, different for each function, namely $Z^2$, $(A-Z)^2$ and $Z(A-Z)$, and the exposure of the detector in the definition of the functions. 

For the SD interactions, one can also perform the same exercise yielding,

\begin{eqnarray}
N_{SD} &= \ \sigma^{SD}_p&\left[\left(1+\left(\frac{a_n}{a_p}\right)^2+2\frac{a_n}{a_p}\right)F^{00}_{Z}(m_{\chi})+\left(1+\left(\frac{a_n}{a_p}\right)^2-2\frac{a_n}{a_p}\right)F^{11}_{Z} (m_{\chi}) \right. \nonumber\\
&& \left.+\left(1-\left(\frac{a_n}{a_p}\right)^2\right)F^{01}_{Z}(m_{\chi})\right],
\label{eq:nsd}
\end{eqnarray}
where the superindices of the $F^{00,01,11}_{Z}$ functions correspond to the isoscalar and isovector decomposition of the SD structure functions. 
These SD functions have been normalized to a reference cross section of $10^{-4}$~pb. 
Therefore, the total number of events in a generic DM experiment can be calculated just by summing the SI and SD contributions. Up to some constants, these functions are the same as those presented in Ref.~\cite{DelNobile:2013sia} for $\mathcal{O}_1$ and $\mathcal{O}_4$ in the framework of the effective field theory for DM interactions.

The maximum gap method finds the most constraining gap between observed events for setting an upper limit. 
This is, the energy range between two observed events with the highest number of events predicted by a model. 
Then, to calculate the bounds we determine the probability of this gap
being smaller than a particular number of events, $x$~\cite{Yellin:2002xd},
\begin{equation}
C_{0}(x,\mu) = \sum_{k=0}^{m} \frac{(kx-\mu)^{k}e^{-kx}}{k!}\left( 1 + \frac{k}{\mu - kx} \right),
\label{eq:C_zero}
\end{equation}
where $\mu$ is the total number of events in the whole search energy window, this is $N_{SI} + N_{SD}$ from Eqs.~\ref{eq:nsi} and \ref{eq:nsd}, and $m$ is the greatest integer $\leq\mu/x$. 
First, note that if there is any observed candidate event for the experiment considered in the search window then $x<\mu$, 
and the functions introduced in Eqs.~\ref{eq:nsi} and \ref{eq:nsd} to calculate $x$ and $\mu$ are not the same, because the energy ranges used in each case are different. 
Second, for zero observed events $\mu=x$ by definition of the maximum gap~\cite{Yellin:2002xd}, and the previous equation reduces to,
\begin{equation}
C_{0}(x) = 1 - e^{-x}.
\label{eq:C_zero_0}
\end{equation}

In the following, we will explain how to use the above mentioned expressions to find the upper limits and 
to test points in the parameter space of a given model,  
for the specific cases of SuperCDMS and LUX.

\noindent $\bullet$ \textbf{SuperCDMS bounds:} Since SuperCDMS observed eleven candidate events in the DM search region~\cite{Agnese:2014aze}, we have to define the functions for different energy ranges. 
All the functions have been extracted accordingly to the experimental setup of Ref.~\cite{Agnese:2014aze} (see Section~\ref{sec:upper} for more details) and following the prescription given in Section~\ref{sec:basics}. 
To evaluate the limits up to DM masses of 30 GeV, two different mass ranges must be considered $m_\chi<5.4$~GeV and $m_\chi\geq5.4$~GeV. 
For each range, there is a different maximum gap and consequently we have to compute a distinct $x$, denoted as $x_1$ and $x_2$.
Now, to calculate $\mu$ and $x_{1,2}$ ($x_1$ or $x_2$, depending on the DM mass), one must use Eqs.~\ref{eq:nsi} and \ref{eq:nsd} with the proper tabulated functions given in the provided datafiles  
(see appendix~\ref{sec:files}).

\noindent To test if a specific particle physics input characterized by $\sigma^{SI}_p$, $f_n/f_p$, $\sigma^{SD}_p$, $a_n/a_p$ and $m_\chi$ is allowed or excluded at e.g. 90\% C.L. by SuperCDMS, 
we compute $\mu$ and $x_{1,2}$ and then evaluate $C_0$ from Eq.~\ref{eq:C_zero}. 
If $C_0\geq0.9$, the input is allowed, otherwise is excluded. On the other side, if we want to determine the upper limit at 90\% C.L. for the SI interactions and $f_n/f_p=1$, 
one has to calculate $\mu$ and $x_{1,2}$ using only the SI functions and choosing $f_n/f_p=1$. 
Afterwards, for a specific DM mass, one finds the  $\sigma^{SI}_p$ such that $C_0=0.90$. 

\noindent $\bullet$ \textbf{LUX bounds:} The calculation of LUX bounds is easier because we have considered zero candidate events. 
In this case, the calculation of $C_0$ only requires the computation of the total number of expected events, $\mu$, for the whole energy range and valid for all DM masses. 
Hence, the suitable expression for $C_0$ is Eq.~\ref{eq:C_zero_0}. 
To derive the necessary functions, we have used the experimental setup described in Ref.~\cite{Akerib:2013tjd} (see Section~\ref{sec:upper} for more details) and 
followed the prescriptions given in Section~\ref{sec:basics}. 
To perform the same tasks as for SuperCDMS, we follow the same procedure, the only difference being the definition of $C_0$.

\section{Attached files}
\label{sec:files}

All the datafiles obtained in this work can be downloaded from the site \url{http://goo.gl/1CDFYi}. 
We provide separated files for the tabulated functions, introduced in Eqs.~\ref{eq:nsi} and \ref{eq:nsd}, corresponding to the SI and SD interactions, 
different speed distributions, and SD structure functions, for DM masses below 200 GeV. 
The files are named according to the following convention:
\begin{center}
SI files: {\tt<Experiment>\char`_SI\char`_<Profile>.txt},
\end{center}
\begin{center}
SD files: {\tt<Experiment>\char`_SD\char`_<Profile>\char`_<FormFactor>.txt},
\end{center}
where {\tt<Experiment> = SuperCDMS}, {\tt LUX}; {\tt<Profile> = SHM}, {\tt NFW}, {\tt Einasto} and {\tt Burkert} for the SHM and the speed distributions of Ref.~\cite{Fornasa:2013iaa}; 
{\tt<FormFactor> = Chiral}, {\tt Ressell} and {\tt Dimitrov} for the SD structure functions of SuperCDMS from Refs.~\cite{Klos:2013rwa,Ressell:1993qm,Dimitrov:1994gc} respectively, 
and {\tt<FormFactor> = Chiral}, {\tt BonnA} and {\tt Nijm} for LUX using results from Refs.~\cite{Klos:2013rwa,Bednyakov:2006ux}.

For LUX, we also give results for DM masses above 200 GeV, for different SD structure functions, and using only the SHM, 
since as we have seen, different speed distributions do not impact on the results for masses above 50 GeV approximately. 
These files contain an extra {\tt\char`_extended} in the filename, but they must be used in the same way as the other files.

In the header of each file, the distribution of the columns is given. E.g. 
in the file {\tt SCDMS\char`_SI\char`_SHM.txt}, we find the tabulated functions needed to calculate $\mu$, $x_1$ and $x_2$ 
(denoted by the suffixes, {\tt\char`_Total}, {\tt\char`_MG1} and {\tt\char`_MG2}, respectively) using Eq.~\ref{eq:nsi},  
taking into account only SI interactions and for the SHM.  
The header of this file reads,

\noindent\small\verb|#######################################################################################|\\
\verb|## Functions for SuperCDMS SI interactions using a SHM halo and the Helm form factor|\\
\verb|## Energy intervals:MG1=[2.411, 2.764]keV,MG2=[3.561, 7.176]keV,Total=[2.0, 12.5]keV|\\
\verb|## Columns:Mass,Fp_MG1,Fp_MG2,Fpn_MG1,Fpn_MG2,Fn_MG1,Fn_MG2,Fp_Total,Fpn_Total,Fn_Total|\\
\verb|#######################################################################################|
\normalsize 

The LUX files only have four columns, for instance the header of the file {\tt LUX\char`_SI\char`_SHM.txt}, containing  
all the necessary information to calculate the SI induced events for a SHM speed distribution reads,

\noindent\small\verb|##############################################################################|\\
\verb|## Functions for LUX SI interactions using a SHM halo and the Helm form factor|\\
\verb|## Energy intervals: Total = MG = [2.000, 30.000]keV|\\
\verb|## Columns: Mass, Fp_Total, Fpn_Total, Fn_Total|\\
\verb|##############################################################################|
\normalsize

Finally, we also provide three example codes for LUX and SuperCDMS, namely {\tt LUXexample.py}, {\tt SCDMSexample.py}, and {\tt LUXexample2.py}. 
These python scripts implement the expressions of appendix~\ref{sec:functions} and illustrate how to use them together with the attached datafiles.
The first two codes evaluate if a point of the parameter space, whose specific characteristics are included in the scripts as an example, is allowed or excluded at 90\% C.L. for each experiment separately. 
The third code reads 1000 inputs from the NMSSM parameter space (not all constraints have been imposed to this data), saved in {\tt testdata.dat}, 
checks if they are excluded or not by LUX, and then plots the allowed and excluded points in the $\sigma^{SI}_{p-\chi}$-$m_\chi$ plane. 
This is useful when one wants to compare a scan over the parameter of a given model with the combined LUX bounds.

%----------------------
%BIBLIOGRAPHY
%----------------------

\end{document}